\documentclass[journal=jpcafh,manuscript=article]{achemso}
\usepackage{longtable}
\usepackage{multirow}
\usepackage{amsmath}
\usepackage{subfigure}
\usepackage{cleveref}
\usepackage[usenames,dvipsnames]{color}
\usepackage{soul}
\usepackage{bm}
\usepackage{rotating}

\usepackage{amssymb}
\usepackage{titlecaps}
\usepackage{xcolor}
\usepackage{soul}
\Addlcwords{a an the of on at with for and in into from by without}
\usepackage{units}
\usepackage{tikz}
\usepackage{import}
\usepackage[normalem]{ulem}
\usepackage[sort&compress,numbers,super]{natbib}
\mciteErrorOnUnknownfalse
\usepackage{xr}

\usepackage{multicol} \usepackage{wrapfig} \usepackage{enumitem}
\usepackage{bm} \usepackage{outlines} 
\SectionNumbersOn

\def\rHH{r_{\mathrm{HH}}} 

\author{Karl P. Horn} \affiliation{Dahlem Center for Complex Quantum
  Systems and Fachbereich Physik, Freie Universit{\"a}t Berlin,
  Arnimallee 14, D-14195 Berlin, Germany} \altaffiliation{These
  authors contributed equally}

\author{Luis Itza Vazquez-Salazar} \affiliation[University of
  Basel]{Department of Chemistry, University of Basel,
  Klingelbergstrasse 80 , CH-4056 Basel, Switzerland.}
\altaffiliation{These authors contributed equally}

\author{Christiane P. Koch} \affiliation{Dahlem Center for Complex
  Quantum Systems and Fachbereich Physik, Freie Universit{\"a}t
  Berlin, Arnimallee 14, D-14195 Berlin, Germany}

\author{Markus Meuwly} \affiliation[University of Basel]{Department of
  Chemistry, University of Basel, Klingelbergstrasse 80 , CH-4056
  Basel, Switzerland.}  \email{m.meuwly@unibas.ch}

\title{Improving Potential Energy Surfaces Using Experimental Feshbach Resonance Tomography}

\begin{document}

\date{\today}

\begin{abstract}
The structure and dynamics of a molecular system is governed by its
potential energy surface (PES), representing the total energy as a
function of the nuclear coordinates. Obtaining accurate potential
energy surfaces is limited by the exponential scaling of Hilbert
space, restricting quantitative predictions of experimental
observables from first principles to small molecules with just a few
electrons. Here, we present an explicitly physics-informed approach
for improving and assessing the quality of families of PESs by
modifying them through linear coordinate transformations based on
experimental data. We demonstrate this ”morphing” of the PES for the
He-H$_{2}^{+}$ complex for reference surfaces at three different levels of
quantum chemistry and using recent comprehensive Feshbach resonance
(FR) measurements. In all cases, the positions and intensities of peaks in the collision cross section are improved. We find these observables to be mainly sensitive to the long-range part of the PES.
\end{abstract}

\section*{Teaser}
We introduce a systematic approach based on experimental data to
validate and improve molecular potential energy surfaces.

\section{Introduction}
The potential energy surface (PES) representing the total energy of a
molecule is a fundamental concept for characterizing the dynamics both
in the gas and condensed phase\cite{wales2018,MM.rev:2023}. With high-quality PESs, the
computation of experimental observables becomes possible with
predictive power at a quantitative level. On the other hand, while all
physical observables depend on it, the PES itself cannot be
observed. This raises the question of how to obtain the most accurate PES
for a given system. From an electronic structure perspective, it is
known that full configuration interaction (FCI) calculations with
large basis sets provide the highest quality for the total energies of a
molecule. However, the unfavourable scaling of FCI with the number of
electrons and basis functions prevents its routine use for
constructing full-dimensional PESs for any molecule consisting of more
than a few light atoms. Alternatively, one may approach the question
from an experimentalist's perspective and argue that the ``most
accurate PES'' is the one that best describes physical
observations. Such an approach has been developed for diatomic
molecules: the rotational Rydberg-Klein-Rees (RKR) method solves the
``inversion problem'' of obtaining the potential energy curve given
spectroscopic information.\cite{child:1988} Rotational RKR has also
been applied to triatomic van der Waals
complexes\cite{nesbitt:1989,nesbitt:1993} but cannot be extended to
molecules of arbitrary size. Indeed, solving the ``inverse problem'',
i.e., determining the PES given experimental observables and an
evolution equation from which these observables are calculated has in
general turned out to be very difficult in chemical
physics.\cite{rabitz:2002} This concerns both the choice of
observables as well as the actual inversion procedure. \\

\noindent
An alternative that is not particularly sensitive to the
dimensionality of the problem is "morphing"
PESs.\cite{MM.morphing:1999,bowman:1991} This method exploits the
topological relationship between a reference and a target
PES. Provided that calculations with the reference PES yield
\textit{qualitatively} correct observables $\mathcal{O_{\rm calc}}$
when compared with experimental observations $\mathcal{O_{\rm exp}}$,
the squared difference $\mathcal{L} = |\mathcal{O_{\rm
    calc}}-\mathcal{O_{\rm exp}}|^2$ can be used to reshape the PES
through linear or non-linear coordinate transformations
("morphing").\cite{MM.morphing:1999} It capitalizes on the correct
overall topology of the reference PES and transmutes it into a new PES
by stretching or compressing internal coordinates and the energy
scale, akin to machining a piece of rubber. Alternatives for reshaping PESs are machine learning-based methods such as
  $\Delta-$ML\cite{rama:2015}, transfer
  learning\cite{smith2019approaching,MM.fad:2022}, or differential
  trajectory reweighting.\cite{thaler:2021} Morphing has
been applied successfully to problems in
spectroscopy,\cite{yurchenko:2016} state-to-state reaction cross
sections,\cite{lorenz:2000} and reaction dynamics\cite{vargas:2019}
for systems with up to 6 atoms\cite{van:2001}. A near-quantitative
reference PES has, however, so far not been available for direct
comparison.  For scattering experiments with He--H$_2^+$ such a PES is
now available.\\

\noindent
The He--H$_2^+$ molecular complex is an ideal proxy for the present work owing to the fact
that the PES can be calculated rigorously at the highest level of
quantum chemistry (FCI). The complex is also interesting in itself,
and the current status of experimental and computational spectroscopy
and reaction dynamics has recently been
reviewed.\cite{heh2.review:2022} He--H$_2^+$, which is isoelectronic
to H$_3$, is stable in its electronic ground state and features a rich
reaction dynamics and spectroscopy. Experimentally, near-dissociative
states\cite{car96:395,gam02:6072} and the low-resolution spectroscopy
were reported for both, He--H$_2^+$ and He--D$_2^+$.\cite{asvany:2021}
Assignments of the vibrational bands were possible by comparing with
bound state calculations utilizing a FCI PES.\cite{MM.heh2:2019} Only
recently, it was possible to estimate the dissociation energy of $\sim
1800$ cm$^{-1}$ from spectroscopic measurements.\cite{asvany:2021}
This compares with earlier bound state calculations using the FCI PES
predicting a value of $D_0 = 1784$ cm$^{-1}$.\cite{MM.heh2:2019} This
value was confirmed from a subsequent focal point analysis resulting
in $D_0 = 1789(4)$ cm$^{-1}$ for para-H$_2^+$.\cite{attila:2022}
Furthermore, a range of reactive collision experiments was carried out
which yielded total and differential cross sections, depending on the
vibrational state of the diatomic,\cite{heh2.review:2022} but with
marked differences between observed and computed results.  In
particular, computationally predicted sharp reactive scattering
resonances have not been found experimentally as of
now.\cite{heh2.review:2022} Finally, the role of nonadiabatic
couplings is of considerable current interest as a means to clarify
the role of geometric phase in reaction outcomes and as a source of
friction in the formation of the He--H$_2^+$ complex in the early
universe. This provides additional impetus for a comprehensive
characterization of this seemingly ``simple'' system.\\

\noindent
The present work uses experimentally measured Feshbach resonances for
He--H$_2^+$\cite{MM.heh2:2023} to morph potential energy surfaces.
Feshbach(-Fano) resonances arise if a bound molecular state on a
potential energy surface of a closed channel couples to scattering
states in an open
channel. \cite{chin2010feshbach,rios2020introduction} The recoil
translational energy is determined from measurements which are
expected to probe large spatial areas of a PES and the underlying
intermolecular interactions.\cite{rios2020introduction} The
redistribution of energy due to the Feshbach resonances has recently
been mapped out comprehensively for He--H$_2^+$ and Ne--H$_2^+$ with
coincidence velocity map imaging of electrons and cations, yielding
very favourable agreement between theory and
experiment.\cite{MM.heh2:2023} In these experiments, the ionic
molecular complexes are generated at separations of up to 10 a$_0$
between the rare gas atom and the molecular ion, confirming that the
experiment indeed probes a large spatial extent of the PES, including
its long-range part.\\

\noindent
Here, morphing is applied to initial PESs ranging from essentially
exact FCI (apart from non-Born-Oppenheimer and remaining basis set
effects) to medium- and lower-level methods, that is, Multi-Reference
Configuration Interaction including the Davidson correction (MRCI+Q)
and second-order M{\o}ller-Plesset perturbation theory (MP2).  This
allows us to determine the sensitivity of the PES and information
content in the experimental observables about local and global
features of the PES and to assess the performance of lower-level
methods (e.g. MP2) compared with FCI.  We found that starting from a
PES of sufficiently high quality, the changes introduced by morphing
can be related to parts of the PES that are probed by the
experiments. At the same time, additional experimental observables,
probing primarily the bound region for He interacting with H$_2^+$,
will be required for morphing at the lower levels of quantum chemical
theory.

\section{Results}
\begin{figure}
\centering
\includegraphics[width=\textwidth,height=0.6\textheight]{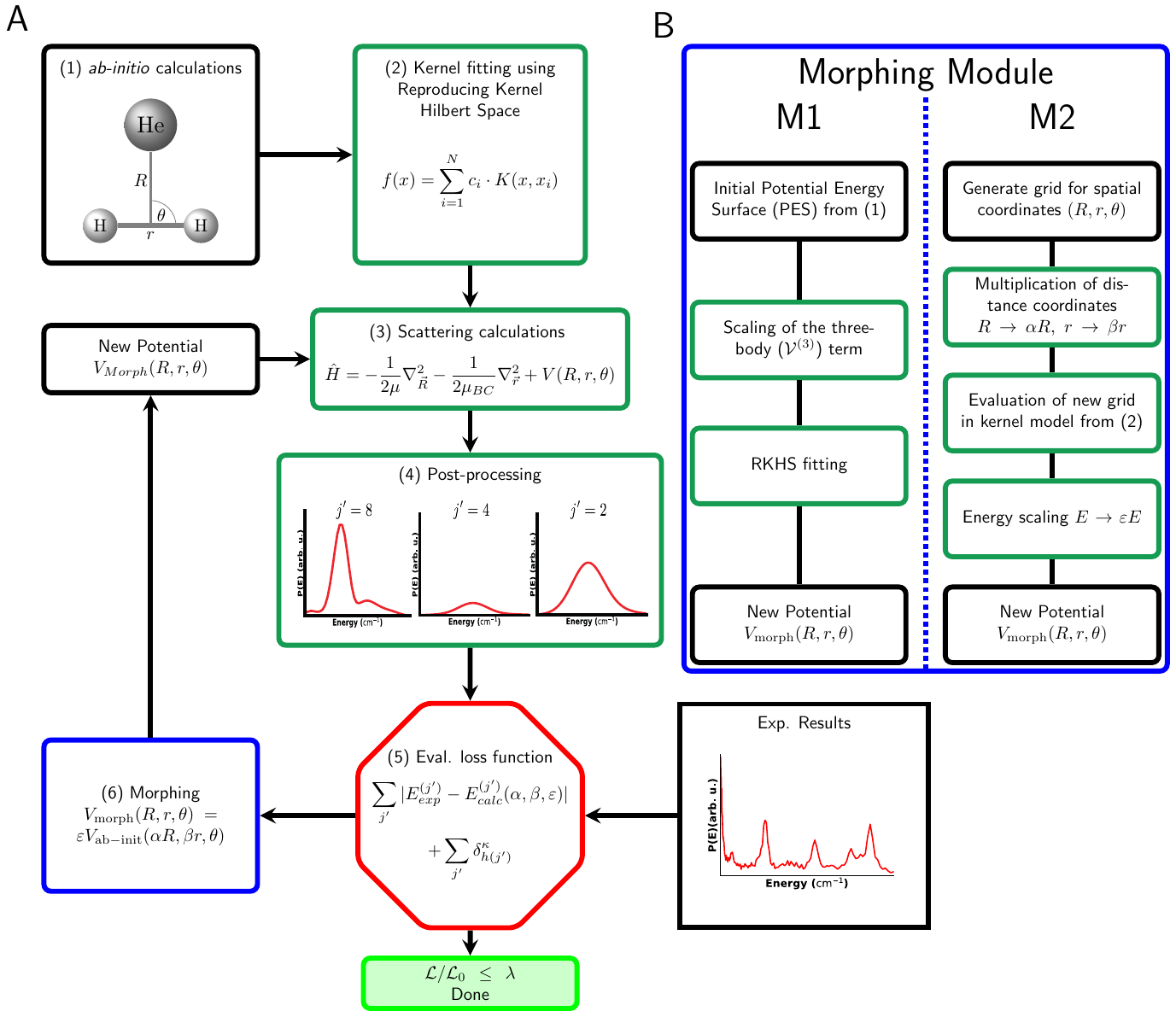}
\caption{Morphing of \textit{ab initio} potentials based on
  experimental data.  General flowchart of the morphing procedure (A):
  Module (1) implements the calculation of \textit{ab-initio} points
  for the system under study, the triatomic HeH$_{2}^{+}$ with the
  definition of the respective coordinates indicated.  Module (2)
  represents the fitting of the points obtained from the previous step
  using the Reproducing Kernel Hilbert Space Method, with the
  functional form used to approximate the given PES.  Module (3)
  corresponds to the scattering calculations performed with the
  potential obtained in module (2), calculating the eigenstates of the
  Hamiltonian. Module (4) post-processes the results of the scattering
  calculations, resulting in the cross sections, with examples for
  three values of $j'$ displayed.  Module (5) evaluates the loss
  function for morphing, comparing the experimental values of the
  cross sections with the results of the scattering
  calculations. Module (6) carries out the actual morphing procedure,
  as explained in panel B. Morphing results in a new potential, and
  the procedure continues until the value of the loss function in
  module (5) does not improve further. The termination
  conditions are $\mathcal{L}/\mathcal{L}_{0}\le \lambda_{\rm M1}=0.3$
  or $\mathcal{L}/\mathcal{L}_{0}\le \lambda_{\rm M2}=0.4$ for M1 and M2, respectively where $\mathcal{L}_0$ is the loss function of the unmorphed cross section, see
  Figure \ref{sifig:error_curve}.    
    Panel B: Morphing
  module (6) for procedures M1 (3-body) and M2 (global).}
\label{fig:arch}
\end{figure}

\begin{figure}[h!]
  \centering
  \includegraphics[scale=0.425]{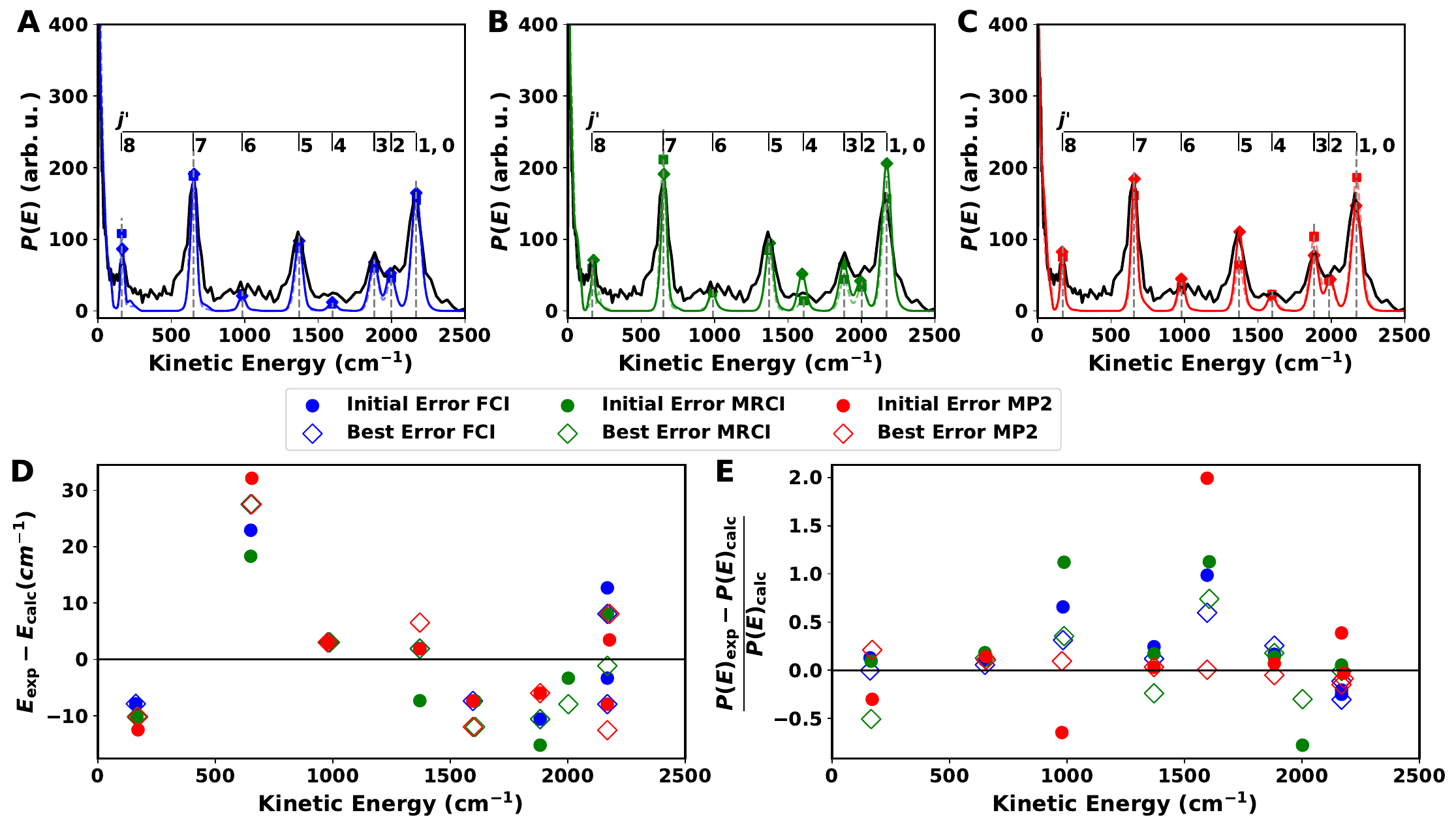}
  \caption{Cross sections obtained from experiment (black, data taken
    from Ref. \citenum{MM.heh2:2023}) and full coupled channels
    calculations using the unmorphed and M1-morphed PESs for FCI (A),
    MRCI+Q (B), and MP2 (C). Computed results for the initial (blue,
    green, red dashed) and best (blue, green, red solid) morphed PESs
    are reported, with the residuals for the peak positions ($E_{\rm
      exp}-E_{\rm calc}$) and fraction of error in the peak heights ($\frac{{P(E)_{\rm exp}-P(E)_{\rm
        calc}}}{P(E)_{\rm
        calc}}$) for each PES shown in Panels D and E. The statistical
    measures for all models are summarized in Table \ref{tab:deeps}.} 
\label{fig:spectra_all}
\end{figure}

The three PESs considered in the present work, in decreasing order of
rigour, were determined at the FCI, MRCI+Q, and MP2 levels of theory,
using Jacobi coordinates $R$ (distance between the centre of mass of
the H$_{2}^{+}$ and He), $r$ (distance between the two hydrogen
atoms), and $\theta$ (the angle between the two vectors $\vec{R}$ and
$\vec{r}$), see Figure \ref{fig:arch}A. To set the stage, scattering
calculations with the FCI PES are considered which give quite
satisfactory results when compared with the experimental data (Figure
\ref{fig:spectra_all} A and Table \ref{tab:deeps}). The measured
kinetic energy distributions feature a series of peaks which reflect
the rovibrational quenching associated with the decay of the Feshbach
resonances.\cite{MM.heh2:2023} On average, the positions of the peak
maxima are reproduced to within 10.8 cm$^{-1}$ whereas the maximum
intensities, $I_{\rm max}$, of $P(E)$ differ by 20.9 arb. u. (blue
squares in Figure \ref{fig:spectra_all}A).\\

\noindent
Next, morphing is applied to all three PESs, including the FCI
PES. The FCI PES has been validated with respect to
experiment\cite{car96:395,gam02:6072,asvany:2021,MM.heh2:2023} and
therefore can serve as a suitable proxy for changes required for PESs
at the MRCI+Q and MP2 levels of theory.  Two morphing strategies were
considered (Figure \ref{fig:arch}B): For Morphing M1, the total energy
was decomposed into one-body ($\mathcal{V}_{i}^{(1)}$), two-body ($\mathcal{V}_{i}^{(2)}$) and three-body($\mathcal{V}^{(3)}$) terms
,
\begin{equation}
    V(R,r,\theta) =\mathcal{V}^{(1)}_{\rm He} + \mathcal{V}^{(1)}_{\rm
      H} + \mathcal{V}^{(1)}_{\rm H^{+}} + \mathcal{V}_{\rm
      HeH}^{(2)}(r_{\rm HeH})+ \mathcal{V}_{\rm HeH^{+}}^{(2)}(r_{\rm
      HeH^{+}}) + \mathcal{V}_{\rm H_{2}^{+}}^{(2)}(r_{\rm H_{2}^{+}})
    + \mathcal{V}^{(3)}(R,r,\theta)\,,
    \label{eq:many-body}
\end{equation}
and the morphing transformation was applied only to the three-body
contribution $\mathcal{V}^{(3)}(R,r,\theta)$. Approach M1 is motivated
by the assumption that all diatomic potentials $\mathcal{V}_{i}^{(2)}$ are of high
quality so that changes are only required in the higher-order
correction three-body term. In the second approach, called ``M2'', the
morphing transformation maps $(R,r) \rightarrow (\alpha R, \beta r)$
and the total energy is multiplied by $\varepsilon$ to obtain the
morphed energy. In this case, the PES is globally modified, including
the two-body contributions.\\

\begin{figure}
\centering
\includegraphics[scale=0.425]{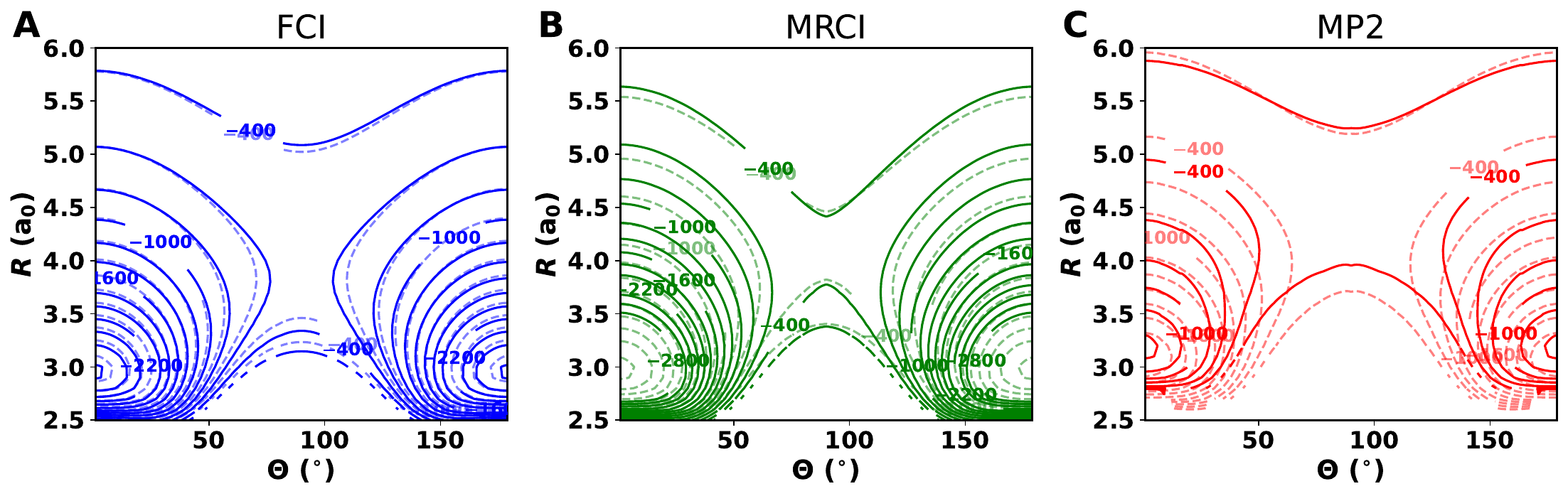}
\caption{Projections of the PESs for $\rHH = 2.0$ a$_{0}$ for the
  three methods studied here. Isocontours for unmorphed PESs (FCI
  (blue), MRCI+Q (green) and MP2 (red) from left to right) are shown
  as dashed lines, whereas the M1-morphed PESs are solid lines. The
  zero of energy is set by the value at $r=2.0$ a$_{0}$ and
  $R=\infty$. Energies are in cm$^{-1}$.}
\label{fig:surfaces_all}
\end{figure}

\noindent
Morphing M1 applied to the FCI PES leaves most of the peak positions
unchanged, see filled vs. open blue symbols in Figure
\ref{fig:spectra_all}D, but improves the peak heights considerably (by
30 \%) as demonstrated in Figure \ref{fig:spectra_all}E and Table
\ref{tab:deeps} (rightmost column).  These improvements are
accommodated by reshaping the underlying PES as shown in Figure
\ref{fig:surfaces_all}A: In the long-range ($R> 3.0$ a$_{0}$), the
anisotropy of the morphed PES is somewhat decreased due to reshaping
the PES around $\theta=90^{\circ}$ (T-shaped geometry). On the other
hand, the curvature along the angular coordinate is
increased. One-dimensional cuts along the $\rHH$ and $R$ coordinates
for a given angle $\theta$ show that changes in the PES become more
significant at larger values of $\rHH$ with small changes in the depth
of the potential wells but maintaining the overall shape of the curves
(Figures \ref{sifig:fci.vdw.1d} and \ref{sifig:fci.r.1d}).  The
changes with respect to $R$ are noticeable for $R<3.0$
a$_{0}$ with small distortions of the energy contours at different angles
$\theta$, but maintaining the overall shape of the curves. For larger
values of $R$, the changes are negligible compared with the original
PES, reflecting the accurate treatment of the long-range interaction
(Figure \ref{sifig:fci.vdw.1d}).  2D projections of the combined
changes of $\rHH$ and $R$ at different angles show that the most
pronounced modifications in the shape of the PES concern regions of
$r_{\rm HH}$ larger than the equilibrium geometry of H$_{2}^{+}$,
i.e., $r_{\rm HH}>2.1$ a$_{0}$, and $R=2-3$ a$_{0}$ (Figures \ref{sifig:Rr_theta0}A
\ref{sifig:Rr_theta30}A and \ref{sifig:Rr_theta60}A).\\

\begin{table}[h!]
\begin{tabular}{l|ccc|cc}
\hline
Surface    & $D_{e}$ (cm$^{-1}$)  & $R_e/$a$_0$    & $r_e/$a$_0$   & RMSE$(E)$  & RMSE$(I)$\\ 
 &   &  &  &  (cm$^{-1}$)   &  (arb. u.)\\ 
\hline \hline
FCI Initial & 2818.9 & 2.97 & 2.07 & 10.8 & 20.9 \\
FCI Morphed (M1)  & 2772.0 & 2.95 & 2.07 & 11.9 & 13.7\\
FCI Morphed (M2)  & 2819.1 & 2.99 & 2.07 & 10.8 & 13.8\\
\hline
MRCI+Q Initial & 2557.3 & 2.98 & 2.07 & 10.3 & 23.9\\
MRCI+Q Morphed (M1) & 3414.7 & 2.98 & 2.08 & 12.2 & 21.9\\
MRCI+Q Morphed (M2) & 2557.0 & 3.00 & 2.03 & 8.9 & 17.6 \\
\hline
MP2 Initial & 2494.0 & 2.99 & 2.07 & 13.1 & 22.4\\
MP2 Morphed (M1)  & 1685.6 & 2.93 & 2.12 & 12.8 & 10.9\\
MP2 Morphed (M2) & 2492.8 & 2.97& 1.74 & 10.0 & 11.8 \\
MP2 Morphed (PES-to-PES) & 2502.3 & 2.98 & 2.06 & 13.0(7) & 22.9 \\  
\hline
\hline
\end{tabular} 
\caption{Dissociation energies ($D_{e}$ in cm$^{-1}$) for He+H$_2^+$,
  coordinates for the minimum energy structures, $R_e$ and $r_e$, and
  root mean squared errors (RMSE) for the peak positions and heights
  of the kinetic energy spectra for all initial and morphed PESs using
  both M1 and M2 methods. In all cases, the equilibrium geometry is
  linear He--H$_2^+$, i.e. $\theta = 0$ or $\theta = 180^\circ$.}
\label{tab:deeps}
\end{table}

\noindent
FCI calculations of entire PESs are only feasible for the smallest
chemical systems, i.e. for diatomic or triatomic molecules. For larger
systems, quantum chemical methods such as multi-reference
configuration interaction or M{\o}ller-Plesset perturbation theory
need to be used instead. As reported in the two rightmost columns of
Table \ref{tab:deeps}, the initial MRCI+Q and MP2 PESs perform almost
on par compared with the initial FCI PES for the positions and
intensities of the kinetic energy distributions. On the other hand,
the dissociation energy is smaller by more than 10\% compared with the
FCI PES due to partial neglect of correlation energy in the MRCI+Q and
MP2 methods. This confirms that Feshbach
resonances are not particularly informative with regards to features
of the PES around the minimum energy structure ($R \sim 3.0$ a$_0$), although the wavefunctions sample this region extensively, see Figure \ref{sifig:wfs}. In other words,
although an important characteristic of a PES such as the stabilization energy of the
complex differs by 10 \% or more, the energies and intensities
measured in collision experiments are matched within similar
bounds.\\

\noindent
Morphing M1 applied to the MRCI+Q and MP2 PESs supports this
observation. The loss function evaluated in module (5) of the
optimization, see Figure \ref{fig:arch}, decreased by 74\% and 88\%
for the two cases, with improvements in the intensities by up to 50\%
for the MP2 PES, see Table \ref{tab:deeps} (rightmost
column). However, the resulting PESs are clearly unphysical, with
pronounced distortions in particular for the MP2 PES, see Figure
\ref{fig:surfaces_all}C and dissociation energies either
increased by 40 \% for MRCI+Q or decreased by 30 \% for MP2,
respectively. Low-resolution experiments\cite{asvany:2021} provide an
estimate for the dissociation energy $D_0 \sim 1800$ cm$^{-1}$,
compared with $D_0 = 1794$ cm$^{-1}$ from bound state calculations on
the initial FCI PES\cite{MM.heh2:2019} which features a well depth of
$D_e \sim 2820$ cm$^{-1}$. This value of $D_e$ serves as a reference
for the remainder of the present work.\\

\noindent
The percentage changes of the parameters $[\alpha,\beta,\varepsilon]$
scaling $(R, r, V)$ provide further information about the
transformation induced by morphing the initial PESs.  For the FCI PES
they are $(-0.6, -3.6, 0.0)\%$ compared with $(-0.6, 11.6, 1.0)$\% for the MRCI+Q and $(0.3, -9.7, 0.1)$\% for the MP2 PES. The most
significant changes concern the H$_2^+$ vibrational coordinate $\rHH$
for MRCI+Q ($+ 12.0$\%) and MP2 ($-10.0$\%).  Such large changes are
problematic since the many-body expansion used for morphing M1,
cf. Eq.~\eqref{eq:many-body}, relies on the quality of the two-body
contributions, i.e., the H$_2^+$ and HeH$^+$ potential energy
curves. However, MP2 underestimates the experimentally determined
dissociation energy of the HeH$^+$ two-body interaction by 285
cm$^{-1}$ (Figure \ref{sifig:2body_hehp}) and accounts for an overall
error of $\sim 500$ cm$^{-1}$ in $D_e$ for He--H$_2^+$. On the other
hand, the two-body term for H$_2^+$ agrees to within 3 cm$^{-1}$
between the three methods with remaining differences compared with the
experiment primarily due to neglect of non-Born-Oppenheimer
contributions (Figure \ref{sifig:2body_h2p}). To summarize: while
M1-morphing improves the match between experimentally measured and
calculated observables, it modifies the PES for the lower-level
methods in an unphysical way. This is attributed to the fact that
M1-morphing operates on the three-body term only and can thus not
compensate for inaccuracies in the two-body contributions to the
overall PES. In contrast, for FCI the changes for all characteristics
of the morphed PES are moderate, underscoring the accuracy of both,
the initial and morphed PESs from FCI calculations.\\

\begin{figure}
    \centering \includegraphics[width=\textwidth]{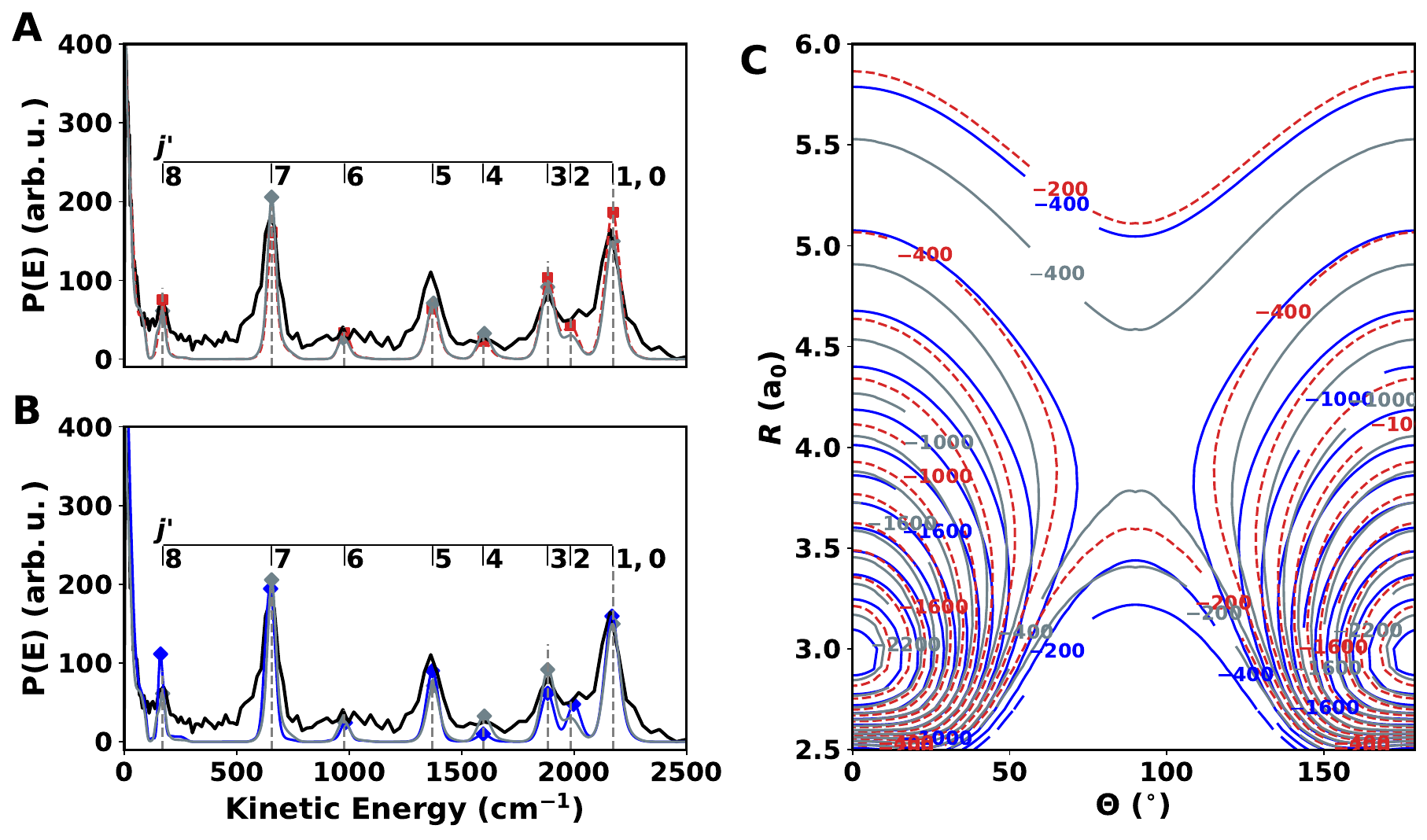}
    \caption{Cross sections (A,C) obtained from experiment (black,
      data taken from Ref. \citenum{MM.heh2:2023}) and full coupled
      channels calculations using the unmorphed (dashed lines) and
      M2-morphed (solid lines) PESs (B,D) for MRCI+Q (A,B), and MP2
      (C,D). The RMSE for the peak positions and heights are reported
      in Table \ref{tab:deeps}. The projections of the PES (B,D) are
      shown for $r=r_{e}$ a$_{0}$ (See Table \ref{tab:deeps}) with the zero of energy set for the
      $r-$value considered and $R=\infty$. Energies are in
      cm$^{-1}$. The changes in the PES suggest that the observables
      are primarily sensitive to the long-range part and the repulsive
      wall of the PES.}
    \label{fig:M2_results}
\end{figure}

\noindent
To reduce the dependence of the morphed PESs on the quality of the
two-body contributions, morphing M2 was carried out. M2-morphing acts
{\it globally} and independently on each of the internal degrees of
freedom, see Figure \ref{fig:arch}. This makes M2 less prone to
overcompensatory effects as observed for M1-morphing. For the MRCI+Q
PES the improvement in the observables amounts to $\approx 14$ \% for
the peak positions and $\approx 26$ \% for the peak heights. At the
same time the changes in the PES are moderate, see Figure
\ref{fig:M2_results}B, and the dissociation energy does not change
(Table \ref{tab:deeps}) although the energy scaling parameter,
$\varepsilon$ was allowed to vary.  Similarly, for MP2, the RMSE for
the positions and heights of the peaks improve by about 22 \% and 47
\%, respectively. Contrary to M1, morphing M2 does not significantly
modify the well depth as reflected by the value of $D_{e}$, see Table
\ref{tab:deeps}.\\

\noindent
For the optimal morphing parameters, M2 applied to the MRCI+Q PES
yields an enlargement of $R$ by $\sim 1$ \% whereas $\rHH$ is reduced
by $1.9\%$ and $\varepsilon$ remains unaffected.  The reduction in
$\rHH$ leads to a small increase in the height of the barrier between
the two wells of the potential (Figure \ref{fig:M2_results}B) and a
corresponding increase in the energy of the transition state, as
observed in the minimum energy path (MEP), see Figure
\ref{sifig:meps_m2}, for the angular coordinate. This effect is
compensated by a positive displacement of the values of $R$ (Figure
\ref{sifig:meps_in_pes_m2}) for the MEP. On the other hand, for the
MP2 surface, the morphing parameters are ($+0.6$, $+ 19.0$, $-0.04$)
\%.  The large positive value for $\beta$ results in a displacement of
the H$_{2}^{+}$ bond length to a shorter equilibrium value (Figure \ref{sifig:Rr_M2} and \ref{sifig:scheme}). For the
$R$ coordinate, the values are also reduced while the barrier height
remains unchanged (Figure \ref{sifig:meps_in_pes_m2}). As for M1, in
the MP2 and MRCI+Q PESs the largest changes are observed in the $\rHH$
coordinate.  However, in the M2 method, scaling of the global PES
results in a better performance for the calculation of the observable
and a better physical description.\\

\begin{figure}
    \centering
    \includegraphics[width=\textwidth]{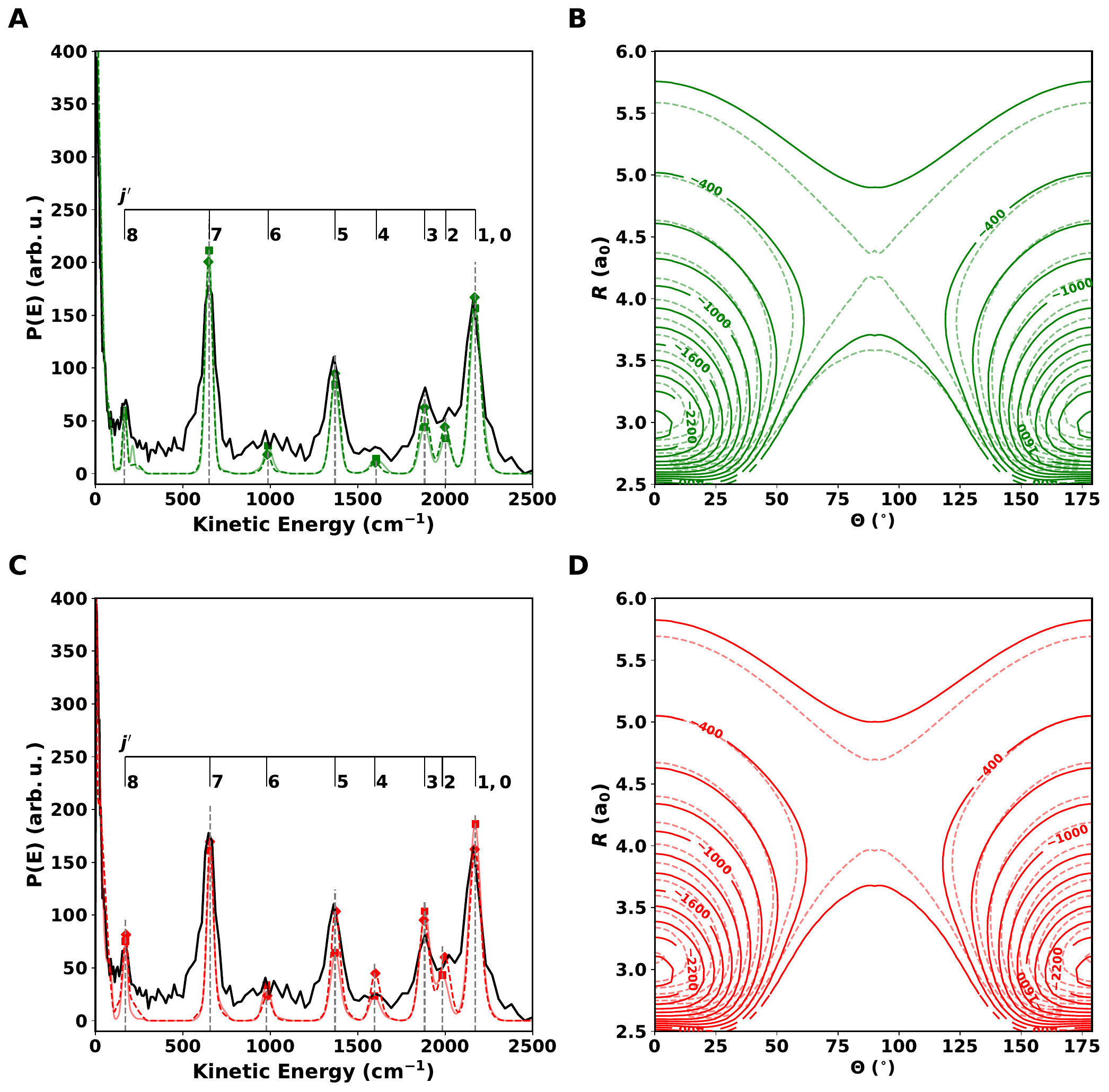}
    \caption{Morphing PES-to-PES. Panel A: Cross-sections obtained
      from experiments (black, data taken from
      Ref. \citenum{MM.heh2:2023}) and scattering calculations on the
      unmorphed MP2 (dashed light red) and the morphed (grey) PESs for
      M2 PES-to-PES morphing procedure with the FCI PES as target.
      Panel B: Same as Panel A but comparing the best morphed PES
      (grey) to the unmorphed FCI surface (solid blue).  Panel C: 2D
      projections of the PES for $r=2.0$ a$_{0}$ for unmorphed FCI
      (solid blue), unmorphed MP2 (dashed light red) and best-morphed
      PES (grey). The zero of energy is set to the value of the PES at
      $\rHH = 2.0$ a$_{0}$ and $R=\infty$. Energies are in cm$^{-1}$.}
    \label{fig:morph_p2p}
\end{figure}

\noindent
Finally, morphing one PES into another one can probe the flexibility
of the morphing transformation as a whole. To this end, the MP2 PES
was morphed to best compare with the FCI PES in a least squares sense
according to method M2, i.e., by finding parameters $(\alpha, \beta,
\varepsilon)$ that minimize $(V_{\rm FCI}(R,r,\theta) - \varepsilon
V_{\rm MP2}(\alpha R, \beta r, \theta))^2$. This optimization
procedure reduces the RMSE between the FCI and unmorphed vs. morphed
PES by about 30\% (from 138 cm$^{-1}$ to 97 cm$^{-1}$, see Figure
\ref{sifig:scatter_pes2pes}). The changes in the topology of the
surface in Figure \ref{fig:morph_p2p}C indicate that the morphed
MP2 PES is "pulled towards" the FCI PES: Consider, for example, the
isocontours for $-400$ cm$^{-1}$ for which the original MP2 isocontour
(blue) is far away from the FCI target contour (red), whereas the
morphed PES (grey) is deformed towards the grey target
isocontour. Closer inspection reveals this to occur for all the other
isocontours in Figure \ref{fig:morph_p2p}C as well. The barrier
separating the [He--HH]$^+$ and [HH--He]$^+$ minima is reduced, which
is also seen in the minimum energy path (see Figure
\ref{sifig:meps_pes2pes}).\\

\noindent
The results of the scattering calculations performed with the surface
from the PES-to-PES morphing procedure (Figure \ref{fig:morph_p2p}A)
are overall slightly inferior to those obtained from the initial FCI
and MP2 PESs, when compared with the experimental data: a negligible
increase of the RMSE for the peak positions ($<1\%$) and intensities
(2.2 \%) is found. Moreover, the fact that the morphing transformation
increases the well depth by merely 10 cm$^{-1}$ indicates that the
simple morphing transformation scaling only distances and the energy
is not sufficiently flexible to accommodate global changes between
topologies as different as FCI vs. MP2.\\

\noindent
The results indicate that at all levels of theory improvements in
describing the experimental observables are possible. At the same time
morphing applied in the fashion done here provides a stringent test to
probe the quality of an initial PES at a quantitative level - with
higher initial levels of theory, the changes that need to be
accommodated decrease and specific deficiencies of a particular
quantum chemical approach can be unveiled.\\

\section{Discussion and Outlook}
Given that essentially exact quantum calculations are possible for the
He--H$_2^+$ complex,\cite{MM.heh2:2019,asvany:2021,MM.heh2:2023} the
present results highlight what can and cannot be learned about
molecular PESs --- the central concept in classical and quantum
molecular dynamics --- from accurate and rather comprehensive
experimental data based on Feshbach resonances. One hallmark of such
quantum scattering resonances is the large spatial extent of the PES
which the resonance wavefunction probes (Figure \ref{sifig:wfs} and discussion in SI). In this regard, the kinetic
energy spectrum obtained from the decay of the Feshbach resonances
differs from spectroscopic observables, typically involving bound
states sensitive to smaller spatial regions of the
PES.\cite{MM.morphing:1999}\\

\noindent
In addition to the actual changes of the PES, a comparison of the two
morphing procedures employed provides insight into the relationship
between the PES, the information provided by specific observables, and
how this information can be used to improve an initial PES.  First,
the much better performance of morphing the global interaction energy
instead of restricting to the three-body contributions reveals the
importance of corrections already at the level of two-body
interactions. Moreover, the physically meaningful changes to the PES
identified by the global morphing concern essentially the anisotropy
in the long range.  To this end, comparatively small changes of the
PESs result in significant improvements in the agreement between
calculated and measured observables.  This is in line with the
expectation that Feshbach resonance wavefunctions mainly probe the
anisotropy of the PES in the long-range.  Both observations taken
together suggest extending the morphing transformation to include
higher-order (and nonlinear) terms as well as the angular degree of
freedom $\theta$ for further improvements.\\

\noindent
At a fundamental level, the present findings raise the question how
much and what experimental data is required to completely characterize
a molecular PES. Indeed, the present work proposes several PESs with
comparable average performance on the scattering observables, even
though the shapes and local characteristics of the PESs differ
greatly, illustrating that the information contained in the Feshbach
resonances is not sufficient to uniquely define the PES. In
particular, information on the bound-state region is missing. One
possible way to answer the question which combination of observables
is suited to completely characterize the dynamics of a molecular
system has been developed in quantum information science and is
referred to as quantum process tomography.\cite{nielsen:2010} It has been adapted to
molecular systems for example in the context of ultrafast
spectroscopy. In future work, quantum process tomography could be
applied to the quest of uniquely defining a PES by making use of the
mapping between the real-space representation of the molecular
Hamiltonian and qubits.\cite{Ollitrault} This should allow for a
systematic approach to identify the most important measurements which
would then provide additional data for morphing PES.\\

\section{Methods}

\subsection{Potential Energy Surfaces}
For the present work, three PESs were employed. Full-dimensional PESs
for He--H$_2^+$ were previously determined at the FCI/aug-cc-pV5Z and
MRCI+Q/aug-cc-pV6Z levels of theory, respectively.\cite{MM.heh2:2019}
The reference data was represented as a reproducing kernel Hilbert
space (RKHS)\cite{MM.rkhs:2017,ho96:2584} which provides a highly
accurate interpolation and allows to encode the leading order
long-range behaviour for large separations. In addition, a third PES
using the same underlying grid for determining reference energies at
the MP2/aug-cc-pV5Z level and also represented as a RKHS, was
constructed for the present work. These calculations were carried out
using the MOLPRO suite of codes.\cite{MOLPRO_brief} All PESs are
represented as a sum of diatomic potential energy curves together with
an explicit three-body interaction.  The complete many-body expansion
for the He--H$_{2}^{+}$ system is given in Eq. (\ref{eq:many-body}),
where distances $r_{i} \in$ \{$r_{\rm HeH},r_{\rm HeH^{+}},r_{\rm
  H_{2}^{+}}\}$ in the two-body terms $\mathcal{V}^{(2)}_{i}$ are the
distances between the respective atoms, whereas for the three-body
term $\mathcal{V}^{(3)}(R,r,\theta)$ the coordinate $r$ is the H$_2^+$
separation $r_{\rm H_{2}^{+}}$, $R$ the distance between He and the
centre of mass of the diatomic, and $\theta$ the angle between the two
distance vectors $\vec{r}$ and $\vec{R}$. Finally, $\mathcal{V}^{(1)}$
corresponds to the respective atomic energies. The energies
$\mathcal{V}_{i}^{(1)}$ and $\mathcal{V}^{(2)}_{i}$ were also determined
at the respective level of theory from electronic structure
calculations and the contributions $\mathcal{V}^{(2)}_{i}$ were fitted
to analytical expressions described in
Ref. \citenum{MM.heh2:2019}. The fitting parameters for the FCI and
MRCI levels of theory were published before and those for the MP2
level of theory are provided in the supporting information. Combining
all this information, the three-body contribution
$\mathcal{V}^{(3)}(R,r,\theta)$ was obtained on the grid used in the
electronic structure calculations for $V(R,r,\theta)$ and represented
as a RKHS.\\

\subsection{Scattering calculations}
Integral scattering cross sections and scattering wave functions for
He-H$_2^+$, resulting from a spatially distributed input wave packet,
were evaluated using a home-written coupled-channels collision
simulation based on the renormalized Numerov
method.\cite{johnson1978renormalized,gadea1997nonradiative} Details on
these calculations have been given in earlier work \cite{MM.heh2:2023}
and only the salient features are presented here. The wavepacket
simulations use Jacobi coordinates with $\vec{r}$ the vector between
the hydrogen atoms, $\vec{R}$ the vector from the dihydrogen centre of
mass to the helium atom and $\theta$ the angle between the two
vectors. With $R=|\vec{R}|$ and $r=|\vec{r}|$, the total Hamiltonian
is then
\begin{equation}
    H_{\mathrm{tot}} = -\frac{\hbar^2}{2\mu_{\mathrm{cmplx}}}
    \nabla^{2}_{\vec{R}} -\frac{\hbar^2}{2\mu_{\mathrm{diat}}}
      \nabla^{2}_{\vec{r}} + V(R,r,\theta)\,\,,
    \label{eq:total_hamiltonian}
\end{equation}
where $\mu_{\mathrm{cmplx}}$ is the reduced mass of the three-body
complex, $\mu_{\mathrm{diat}}$ the reduced mass of the dihydrogen
molecule, and $V(R,r,\theta)$ the three-dimensional PES. The total
wavefunction of the system $\Psi(\vec{R},\vec{r})$ is written as a
product of $R-$, $r-$, and angularly dependent terms,
\begin{equation}
    \Psi^{JMvj\ell}(\vec{R},\vec{r}) \propto
    \sum_{v'j'\ell'} G_{v'j'\ell'}^{Jvj\ell}(R)
    \chi_{\mathrm{diat},v'j'}(r)
    \sum_{m_j=-j}^{j}\sum_{m_{\ell}=-\ell}^{\ell}
    C_{m_j m_{\ell}}^{JM}Y_{\ell,m_{\ell}}(\theta_{R},\varphi_{R})Y_{j,m_j}(\theta_r,\varphi_{r})\,\,,
  \label{eq:wavefunction_expansion}
\end{equation}
see Ref.\citenum{MM.heh2:2023} for more detail. Channels consist of
tuples of quantum numbers $v, j$, and $\ell$, corresponding to
diatomic vibration, rotation and orbital angular momentum,
respectively.  In Eq. (\ref{eq:wavefunction_expansion}),
$\chi_{\mathrm{diat},v,j}(r)$ designates the rovibrational eigenstates
of the molecule.  Starting from a given entrance channel, the
Schr\"odinger equation is solved numerically to obtain the radial wave
functions $G(R)$ for the exit channel with quantum numbers
$(v',j',\ell')$ connected with the entrance channel $(v,j,\ell)$. The
total angular momentum, $\vec{J}_{\mathrm{tot}} = \vec{j}+\vec{L}$
obtained from coupling diatomic and orbital rotation, and parity are
conserved under the Hamiltonian (\ref{eq:total_hamiltonian}).\\

\noindent
In the experiments, the He--H$_2^+$ complex (plus a leaving electron)
is formed by Penning ionization (He$^*$+H$_2$), and the scattering
calculations considered in the present work describe the
half-collision on the He--H$_2^+$ PES.  The initial wavepacket
$\phi(R)$ along the $R-$coordinate is approximated by Gaussian
distributions centered around $R \approx 8~a_{0}$.\cite{MM.heh2:2023}
The experiment prepares the input wavepacket with $j_{\mathrm{wp}}=0,
1$ for para- and ortho-H$_2^+$, respectively. However, as the system
is prepared in a superposition of $J-$states, individual simulations
need to be carried out for each possible value of $J$ and partial wave
$\ell$.  Then, the integral cross section is calculated as a weighted
sum over the individual contributions for a given collision energy
$E_{\mathrm{col}}/ \mathrm{k_{B}} \approx 2.5$ K. The $J-$weights,
which were calculated separately,\cite{pawlak2017adiabatic} are shown
in Figure \ref{sifig:weights}.\\

\noindent
Evaluation of the collision cross section due to the spatially
distributed input wavepacket can be accomplished by expanding
$\phi(R)$ in a basis of eigenfunctions of $H_{\mathrm{tot}}$. To this
end, the time-independent Schr{\"o}dinger equation was solved on a
discretized interval
of 1002 energies ranging from $100~\mathrm{cm}^{-1}$
below to $100~\mathrm{cm}^{-1}$ above the dissociation threshold of
the given entrance channel. Because full coupled-channel calculations
are computationally demanding, the considered set of initial
wavepacket quantum numbers $J$ and $\ell$ was limited to $(\ell/J) \in
\{(0/0),(1/1),(2/2),(3/3),(4/4)\}$ for para- and $(\ell/J) \in
\{(0/1),(1/1,2),(2/1, 2, 3)$ $,(3/2, 3, 4),(4/3, 4, 5)\}$ for
ortho-dihydrogen, respectively. For each coupled channel calculation a
converged basis set of diatomic rotational states up to $j_{\rm
  max}=19$ and diatomic vibrational states up to $v_{\max}=5$ was
used.\\

\noindent
Solving the Schr{\"o}dinger equation in this fashion allows for
calculating the channel-resolved integral cross section for each
energy in the discretized interval.  For a given output channel, the
eigenenergy $E_{v'j'\ell'}=E_{\mathrm{int},v'j'\ell'} +
E_{\mathrm{kin},v',j',\ell'}$ can be decomposed into its internal and
kinetic parts, respectively.  By generating a histogram for all output
channels ($v'$,$j'$,$\ell'$), the cross-section can be expressed as a
function of kinetic energy, which can be compared with the
experimental results. Next, the kinetic energy histogram is convoluted
using a Gaussian envelope to account for the finite resolution in the
experiments.\cite{MM.heh2:2023} Before convolution, and as shown in
Figure \ref{sifig:vwp1theo}, the computed peaks are sharp in
$E_{\mathrm{kin}}$ which is a signature of Feshbach resonances. It
should be noted that experimental peaks are clearly distinguishable
and energetically match the theoretical predictions. However, the peak
shapes and heights can vary, dependent on the histogram resolution and
convolution width.  In this work, only single initial vibrational
excitations ($v=1$) were considered, in order to exploit the
experimental resolution of separate $j'$ peaks in the cross-section as
a function of kinetic energy \cite{margulis2020direct}.\\

\subsection{Morphing}
The morphing transformation considered here is
\begin{equation}
V_{\mathrm{morphed}}(R,r,\theta) = \varepsilon
V_{\mathrm{ab-initio}}(\alpha R, \beta r, \theta)\,\,.
\label{eq:morph_used}
\end{equation}
In Eq. (\ref{eq:morph_used}), the three parameters
($\alpha,\beta,\varepsilon$) are used for energy- ($\varepsilon$) and
geometry-($\alpha, \beta$) related scalings. For the purpose of this
work, the angle $\theta$ was not modified. The morphing procedure
described further below optimizes the values of $(\alpha, \beta,
\varepsilon)$ such that the difference between observed and computed
features of the resonances is minimized. Application of such a
procedure modifies local features (e.g. slope, curvature) of the PES
but maintains its global shape.\\

\noindent
For morphing M1 and M2 the refinement with respect to experimental
values is formulated as an optimization problem with a loss function,
\begin{equation}
\mathcal{L} = \min_{\alpha,\beta,\varepsilon} \left[ \sum_{j'}
  |E_{\mathrm{exp}}^{(j')}-E_{\mathrm{calc}}^{(j')}(\alpha,\beta,\varepsilon)|
  + \sum_{j'} \delta_{h(j')}^{\kappa} \right]\,\,,
\label{eq:loss}
\end{equation}
to be minimized. Here, $E^{(j')}$ is the kinetic energy of each
cross-section corresponding to an exit-channel $j'$, and $\delta_{h(j')}^{\kappa}$
accounts for the difference in the peak heights
between experimental and calculated values:
\begin{equation}
    \delta_{h(j')}^{\kappa} = \left\{
    \begin{array}{cr}
        (\Delta h(j')
        - h_{\mathrm{noise}})^{\kappa},  & (\Delta h(j')
        - h_{\mathrm{noise}})^{\kappa} > 0\,\,,\\
        0,  & (\Delta h(j')
        - h_{\mathrm{noise}})^{\kappa} \leq 0\,\,,
    \end{array} \right. 
    \label{eq:regularized_peak_heights}
\end{equation}
\noindent
where, $\delta_{h} (j')$ is \textit{regularized} by subtracting
$h_{\mathrm{noise}} =10.0$ to avoid fitting experimental noise. By
design, only values $\delta_{h} (j') > 0$ contribute to the error.
Here $\Delta h(j')=|h^{(j')}_{\mathrm{exp}} - \gamma
h^{(j')}_{\mathrm{calc}}(\alpha,\beta,\varepsilon)|$, where $h^{(j')}$
is the peak height of the cross section corresponding to an
exit-channel $j'$. The parameter $\gamma$ is recalculated after each
iteration to best match the experiment by performing an additional 1d
minimization over the squared difference in peaks heights.\\

\noindent
The workflow to perform the optimization of Eq. (\ref{eq:loss}) is
shown schematically in Figure \ref{fig:arch}. In the first step, {\it
  ab initio} points of the PES are used to generate a RKHS
kernel. Depending on the morphing procedure chosen, a new RKHS needs
to be generated (for M1) or the existing kernel will be reused (for
M2). All kernels are constructed and evaluated using the ``fast''
method.\cite{MM.rkhs:2017} The obtained PES is passed to the
scattering code to perform the wavepacket propagation. Next, the
resulting cross-sections are processed and then compared with the
available experimental data. If the difference between experimental
and calculated values matches a given tolerance the cycle finishes;
otherwise, the PES is modified by three parameters as described in
Eq. (\ref{eq:morph_used}) following the chosen morphing approach.  The
values of the parameters $\alpha$, $\beta$ and $\varepsilon$ were
obtained by a non-linear optimization using the NLopt
package\cite{nlopt}. \\

\section*{Data Availability Statement}
The data needed for the PESs is available at
\url{https://github.com/MeuwlyGroup/morphing}.

\section*{Acknowledgments}
This work was supported by the Swiss National Science Foundation
grants 200021-117810, 200020-188724, the NCCR MUST, and the University
of Basel which is gratefully acknowledged. The authors thank
B. Margulis and Prof. E. Narevicius for providing the experimental
data used in this work.

\bibliography{refs.bib}

\providecommand{\latin}[1]{#1}
\makeatletter
\providecommand{\doi}
  {\begingroup\let\do\@makeother\dospecials
  \catcode`\{=1 \catcode`\}=2 \doi@aux}
\providecommand{\doi@aux}[1]{\endgroup\texttt{#1}}
\makeatother
\providecommand*\mcitethebibliography{\thebibliography}
\csname @ifundefined\endcsname{endmcitethebibliography}
  {\let\endmcitethebibliography\endthebibliography}{}
\begin{mcitethebibliography}{45}
\providecommand*\natexlab[1]{#1}
\providecommand*\mciteSetBstSublistMode[1]{}
\providecommand*\mciteSetBstMaxWidthForm[2]{}
\providecommand*\mciteBstWouldAddEndPuncttrue
  {\def\EndOfBibitem{\unskip.}}
\providecommand*\mciteBstWouldAddEndPunctfalse
  {\let\EndOfBibitem\relax}
\providecommand*\mciteSetBstMidEndSepPunct[3]{}
\providecommand*\mciteSetBstSublistLabelBeginEnd[3]{}
\providecommand*\EndOfBibitem{}
\mciteSetBstSublistMode{f}
\mciteSetBstMaxWidthForm{subitem}{(\alph{mcitesubitemcount})}
\mciteSetBstSublistLabelBeginEnd
  {\mcitemaxwidthsubitemform\space}
  {\relax}
  {\relax}

\bibitem[Wales(2018)]{wales2018}
Wales,~D.~J. Exploring Energy Landscapes. \emph{Annu.\ Rev.\ Phys.\ Chem.}
  \textbf{2018}, \emph{69}, 401--425\relax
\mciteBstWouldAddEndPuncttrue
\mciteSetBstMidEndSepPunct{\mcitedefaultmidpunct}
{\mcitedefaultendpunct}{\mcitedefaultseppunct}\relax
\EndOfBibitem
\bibitem[K{\"a}ser \latin{et~al.}(2023)K{\"a}ser, Vazquez-Salazar, Meuwly, and
  T{\"o}pfer]{MM.rev:2023}
K{\"a}ser,~S.; Vazquez-Salazar,~L.~I.; Meuwly,~M.; T{\"o}pfer,~K. Neural
  network potentials for chemistry: concepts, applications and prospects.
  \emph{Dig. Discov.} \textbf{2023}, \emph{2}, 28--58\relax
\mciteBstWouldAddEndPuncttrue
\mciteSetBstMidEndSepPunct{\mcitedefaultmidpunct}
{\mcitedefaultendpunct}{\mcitedefaultseppunct}\relax
\EndOfBibitem
\bibitem[Child and Nesbitt(1988)Child, and Nesbitt]{child:1988}
Child,~M.; Nesbitt,~D. RKR-based inversion of rotational progressions.
  \emph{Chem.\ Phys.\ Lett.} \textbf{1988}, \emph{149}, 404--410\relax
\mciteBstWouldAddEndPuncttrue
\mciteSetBstMidEndSepPunct{\mcitedefaultmidpunct}
{\mcitedefaultendpunct}{\mcitedefaultseppunct}\relax
\EndOfBibitem
\bibitem[Nesbitt \latin{et~al.}(1989)Nesbitt, Child, and Clary]{nesbitt:1989}
Nesbitt,~D.~J.; Child,~M.~S.; Clary,~D.~C. Rydberg--Klein--Rees inversion of
  high resolution van der Waals infrared spectra: An intermolecular potential
  energy surface for Ar+ HF$(v= 1)$. \emph{J.~Chem.\ Phys.} \textbf{1989},
  \emph{90}, 4855--4864\relax
\mciteBstWouldAddEndPuncttrue
\mciteSetBstMidEndSepPunct{\mcitedefaultmidpunct}
{\mcitedefaultendpunct}{\mcitedefaultseppunct}\relax
\EndOfBibitem
\bibitem[Nesbitt and Child(1993)Nesbitt, and Child]{nesbitt:1993}
Nesbitt,~D.~J.; Child,~M.~S. Rotational-RKR inversion of intermolecular
  stretching potentials: Extension to linear hydrogen bonded complexes.
  \emph{J.~Chem.\ Phys.} \textbf{1993}, \emph{98}, 478--486\relax
\mciteBstWouldAddEndPuncttrue
\mciteSetBstMidEndSepPunct{\mcitedefaultmidpunct}
{\mcitedefaultendpunct}{\mcitedefaultseppunct}\relax
\EndOfBibitem
\bibitem[Kurtz \latin{et~al.}(2002)Kurtz, Rabitz, and
  de~Vivie-Riedle]{rabitz:2002}
Kurtz,~L.; Rabitz,~H.; de~Vivie-Riedle,~R. Optimal use of time-dependent
  probability density data to extract potential-energy surfaces. \emph{Phys.\
  Rev.~A} \textbf{2002}, \emph{65}, 032514\relax
\mciteBstWouldAddEndPuncttrue
\mciteSetBstMidEndSepPunct{\mcitedefaultmidpunct}
{\mcitedefaultendpunct}{\mcitedefaultseppunct}\relax
\EndOfBibitem
\bibitem[Meuwly and Hutson(1999)Meuwly, and Hutson]{MM.morphing:1999}
Meuwly,~M.; Hutson,~J.~M. Morphing ab initio potentials: A systematic study of
  Ne--HF. \emph{J.~Chem.\ Phys.} \textbf{1999}, \emph{110}, 8338--8347\relax
\mciteBstWouldAddEndPuncttrue
\mciteSetBstMidEndSepPunct{\mcitedefaultmidpunct}
{\mcitedefaultendpunct}{\mcitedefaultseppunct}\relax
\EndOfBibitem
\bibitem[Bowman and Gazdy(1991)Bowman, and Gazdy]{bowman:1991}
Bowman,~J.~M.; Gazdy,~B. A simple method to adjust potential energy surfaces:
  Application to HCO. \emph{J.~Chem.\ Phys.} \textbf{1991}, \emph{94},
  816--817\relax
\mciteBstWouldAddEndPuncttrue
\mciteSetBstMidEndSepPunct{\mcitedefaultmidpunct}
{\mcitedefaultendpunct}{\mcitedefaultseppunct}\relax
\EndOfBibitem
\bibitem[Ramakrishnan \latin{et~al.}(2015)Ramakrishnan, Dral, Rupp, and
  Von~Lilienfeld]{rama:2015}
Ramakrishnan,~R.; Dral,~P.~O.; Rupp,~M.; Von~Lilienfeld,~O.~A. Big data meets
  quantum chemistry approximations: the $\Delta$-machine learning approach.
  \emph{J.~Chem.\ Theor.\ Comp.} \textbf{2015}, \emph{11}, 2087--2096\relax
\mciteBstWouldAddEndPuncttrue
\mciteSetBstMidEndSepPunct{\mcitedefaultmidpunct}
{\mcitedefaultendpunct}{\mcitedefaultseppunct}\relax
\EndOfBibitem
\bibitem[Smith \latin{et~al.}(2019)Smith, Nebgen, Zubatyuk, Lubbers, Devereux,
  Barros, Tretiak, Isayev, and Roitberg]{smith2019approaching}
Smith,~J.~S.; Nebgen,~B.~T.; Zubatyuk,~R.; Lubbers,~N.; Devereux,~C.;
  Barros,~K.; Tretiak,~S.; Isayev,~O.; Roitberg,~A.~E. Approaching coupled
  cluster accuracy with a general-purpose neural network potential through
  transfer learning. \emph{Nat. Commun.} \textbf{2019}, \emph{10}, 1--8\relax
\mciteBstWouldAddEndPuncttrue
\mciteSetBstMidEndSepPunct{\mcitedefaultmidpunct}
{\mcitedefaultendpunct}{\mcitedefaultseppunct}\relax
\EndOfBibitem
\bibitem[K{\"a}ser and Meuwly(2022)K{\"a}ser, and Meuwly]{MM.fad:2022}
K{\"a}ser,~S.; Meuwly,~M. Transfer learned potential energy surfaces: accurate
  anharmonic vibrational dynamics and dissociation energies for the formic acid
  monomer and dimer. \emph{Phys.\ Chem.\ Chem.\ Phys.} \textbf{2022},
  \emph{24}, 5269--5281\relax
\mciteBstWouldAddEndPuncttrue
\mciteSetBstMidEndSepPunct{\mcitedefaultmidpunct}
{\mcitedefaultendpunct}{\mcitedefaultseppunct}\relax
\EndOfBibitem
\bibitem[Thaler and Zavadlav(2021)Thaler, and Zavadlav]{thaler:2021}
Thaler,~S.; Zavadlav,~J. Learning neural network potentials from experimental
  data via Differentiable Trajectory Reweighting. \emph{Nat. Comm.}
  \textbf{2021}, \emph{12}, 6884\relax
\mciteBstWouldAddEndPuncttrue
\mciteSetBstMidEndSepPunct{\mcitedefaultmidpunct}
{\mcitedefaultendpunct}{\mcitedefaultseppunct}\relax
\EndOfBibitem
\bibitem[Yurchenko \latin{et~al.}(2016)Yurchenko, Lodi, Tennyson, and
  Stolyarov]{yurchenko:2016}
Yurchenko,~S.~N.; Lodi,~L.; Tennyson,~J.; Stolyarov,~A.~V. Duo: A general
  program for calculating spectra of diatomic molecules. \emph{Comput.\ Phys.\
  Commun.} \textbf{2016}, \emph{202}, 262--275\relax
\mciteBstWouldAddEndPuncttrue
\mciteSetBstMidEndSepPunct{\mcitedefaultmidpunct}
{\mcitedefaultendpunct}{\mcitedefaultseppunct}\relax
\EndOfBibitem
\bibitem[Lorenz \latin{et~al.}(2000)Lorenz, Westley, and Chandler]{lorenz:2000}
Lorenz,~K.~T.; Westley,~M.~S.; Chandler,~D.~W. Rotational state-to-state
  differential cross sections for the HCl--Ar collision system using
  velocity-mapped ion imaging. \emph{Phys.\ Chem.\ Chem.\ Phys.} \textbf{2000},
  \emph{2}, 481--494\relax
\mciteBstWouldAddEndPuncttrue
\mciteSetBstMidEndSepPunct{\mcitedefaultmidpunct}
{\mcitedefaultendpunct}{\mcitedefaultseppunct}\relax
\EndOfBibitem
\bibitem[Vargas-Hern{\'a}ndez \latin{et~al.}(2019)Vargas-Hern{\'a}ndez, Guan,
  Zhang, and Krems]{vargas:2019}
Vargas-Hern{\'a}ndez,~R.; Guan,~Y.; Zhang,~D.; Krems,~R. Bayesian optimization
  for the inverse scattering problem in quantum reaction dynamics. \emph{New J.
  Phys.} \textbf{2019}, \emph{21}, 022001\relax
\mciteBstWouldAddEndPuncttrue
\mciteSetBstMidEndSepPunct{\mcitedefaultmidpunct}
{\mcitedefaultendpunct}{\mcitedefaultseppunct}\relax
\EndOfBibitem
\bibitem[van Mourik \latin{et~al.}(2001)van Mourik, Harris, Polyansky,
  Tennyson, Cs{\'a}sz{\'a}r, and Knowles]{van:2001}
van Mourik,~T.; Harris,~G.~J.; Polyansky,~O.~L.; Tennyson,~J.;
  Cs{\'a}sz{\'a}r,~A.~G.; Knowles,~P.~J. Ab initio global potential, dipole,
  adiabatic, and relativistic correction surfaces for the HCN--HNC system.
  \emph{J.~Chem.\ Phys.} \textbf{2001}, \emph{115}, 3706--3718\relax
\mciteBstWouldAddEndPuncttrue
\mciteSetBstMidEndSepPunct{\mcitedefaultmidpunct}
{\mcitedefaultendpunct}{\mcitedefaultseppunct}\relax
\EndOfBibitem
\bibitem[Adhikari \latin{et~al.}(2022)Adhikari, Baer, and
  Sathyamurthy]{heh2.review:2022}
Adhikari,~S.; Baer,~M.; Sathyamurthy,~N. HeH$_{2}^{+}$: structure and dynamics.
  \emph{Int.\ Rev.\ Phys.\ Chem.} \textbf{2022}, \emph{41}, 49--93\relax
\mciteBstWouldAddEndPuncttrue
\mciteSetBstMidEndSepPunct{\mcitedefaultmidpunct}
{\mcitedefaultendpunct}{\mcitedefaultseppunct}\relax
\EndOfBibitem
\bibitem[Carrington \latin{et~al.}(1996)Carrington, Gammie, Shaw, Taylor, and
  Hutson]{car96:395}
Carrington,~A.; Gammie,~D.~I.; Shaw,~A.~M.; Taylor,~S.~M.; Hutson,~J.~M.
  {Observation of a microwave spectrum of the long-range He $\ldots$
  H$_{2}^{+}$ complex}. \emph{Chem. Phys. Lett.} \textbf{1996}, \emph{260},
  395--405\relax
\mciteBstWouldAddEndPuncttrue
\mciteSetBstMidEndSepPunct{\mcitedefaultmidpunct}
{\mcitedefaultendpunct}{\mcitedefaultseppunct}\relax
\EndOfBibitem
\bibitem[Gammie \latin{et~al.}(2002)Gammie, Page, and Shaw]{gam02:6072}
Gammie,~D.~I.; Page,~J.~C.; Shaw,~A.~M. {Microwave and millimeter-wave spectrum
  of the He$\cdots$H$^{+}_{2}$ long-range complex}. \emph{J.~Chem.\ Phys.}
  \textbf{2002}, \emph{116}, 6072\relax
\mciteBstWouldAddEndPuncttrue
\mciteSetBstMidEndSepPunct{\mcitedefaultmidpunct}
{\mcitedefaultendpunct}{\mcitedefaultseppunct}\relax
\EndOfBibitem
\bibitem[Asvany \latin{et~al.}(2021)Asvany, Schlemmer, van~der Avoird,
  Szidarovszky, and Cs{\'a}sz{\'a}r]{asvany:2021}
Asvany,~O.; Schlemmer,~S.; van~der Avoird,~A.; Szidarovszky,~T.;
  Cs{\'a}sz{\'a}r,~A.~G. Vibrational spectroscopy of H$_2$He$^+$ and
  D$_2$He$^+$. \emph{J.\ Mol.\ Struct.} \textbf{2021}, \emph{377}, 111423\relax
\mciteBstWouldAddEndPuncttrue
\mciteSetBstMidEndSepPunct{\mcitedefaultmidpunct}
{\mcitedefaultendpunct}{\mcitedefaultseppunct}\relax
\EndOfBibitem
\bibitem[Koner \latin{et~al.}(2019)Koner, Veliz, van~der Avoird, and
  Meuwly]{MM.heh2:2019}
Koner,~D.; Veliz,~J. C. S.~V.; van~der Avoird,~A.; Meuwly,~M. Near dissociation
  states for H$_2^+$--He on MRCI and FCI potential energy surfaces.
  \emph{Phys.\ Chem.\ Chem.\ Phys.} \textbf{2019}, \emph{21},
  24976--24983\relax
\mciteBstWouldAddEndPuncttrue
\mciteSetBstMidEndSepPunct{\mcitedefaultmidpunct}
{\mcitedefaultendpunct}{\mcitedefaultseppunct}\relax
\EndOfBibitem
\bibitem[Kedziera \latin{et~al.}(2022)Kedziera, Rauhut, and
  Cs{\'a}sz{\'a}r]{attila:2022}
Kedziera,~D.; Rauhut,~G.; Cs{\'a}sz{\'a}r,~A.~G. Structure, energetics, and
  spectroscopy of the chromophores of HHe$_n^+$, H$_2$He$_n^+$, and He$_n^+$
  clusters and their deuterated isotopologues. \emph{Phys.\ Chem.\ Chem.\
  Phys.} \textbf{2022}, \emph{24}, 12176--12195\relax
\mciteBstWouldAddEndPuncttrue
\mciteSetBstMidEndSepPunct{\mcitedefaultmidpunct}
{\mcitedefaultendpunct}{\mcitedefaultseppunct}\relax
\EndOfBibitem
\bibitem[Margulis \latin{et~al.}(2023)Margulis, Horn, Reich, Upadhyay, Kahn,
  Christianen, van~der Avoird, Groenenboom, Koch, Meuwly, and
  Narevicius]{MM.heh2:2023}
Margulis,~B.; Horn,~K.~P.; Reich,~D.~M.; Upadhyay,~M.; Kahn,~N.;
  Christianen,~A.; van~der Avoird,~A.; Groenenboom,~G.~C.; Koch,~C.~P.;
  Meuwly,~M. \latin{et~al.}  Tomography of Feshbach resonance states.
  \emph{Science} \textbf{2023}, \emph{380}, 77--81\relax
\mciteBstWouldAddEndPuncttrue
\mciteSetBstMidEndSepPunct{\mcitedefaultmidpunct}
{\mcitedefaultendpunct}{\mcitedefaultseppunct}\relax
\EndOfBibitem
\bibitem[Chin \latin{et~al.}(2010)Chin, Grimm, Julienne, and
  Tiesinga]{chin2010feshbach}
Chin,~C.; Grimm,~R.; Julienne,~P.; Tiesinga,~E. Feshbach resonances in
  ultracold gases. \emph{Rev.\ Mod.\ Phys.} \textbf{2010}, \emph{82},
  1225\relax
\mciteBstWouldAddEndPuncttrue
\mciteSetBstMidEndSepPunct{\mcitedefaultmidpunct}
{\mcitedefaultendpunct}{\mcitedefaultseppunct}\relax
\EndOfBibitem
\bibitem[P{\'e}rez~R{\'i}os(2020)]{rios2020introduction}
P{\'e}rez~R{\'i}os,~J. \emph{Introduction to Cold and Ultracold Chemistry:
  Atoms, Molecules, Ions and Rydbergs}; Springer, 2020\relax
\mciteBstWouldAddEndPuncttrue
\mciteSetBstMidEndSepPunct{\mcitedefaultmidpunct}
{\mcitedefaultendpunct}{\mcitedefaultseppunct}\relax
\EndOfBibitem
\bibitem[Nielsen and Chuang(2010)Nielsen, and Chuang]{nielsen:2010}
Nielsen,~M.~A.; Chuang,~I.~L. \emph{Quantum computation and quantum
  information}; Cambridge university press, 2010\relax
\mciteBstWouldAddEndPuncttrue
\mciteSetBstMidEndSepPunct{\mcitedefaultmidpunct}
{\mcitedefaultendpunct}{\mcitedefaultseppunct}\relax
\EndOfBibitem
\bibitem[Ollitrault \latin{et~al.}(2021)Ollitrault, Miessen, and
  Tavernelli]{Ollitrault}
Ollitrault,~P.~J.; Miessen,~A.; Tavernelli,~I. Molecular Quantum Dynamics: A
  Quantum Computing Perspective. \emph{Accounts of Chemical Research}
  \textbf{2021}, \emph{54}, 4229--4238\relax
\mciteBstWouldAddEndPuncttrue
\mciteSetBstMidEndSepPunct{\mcitedefaultmidpunct}
{\mcitedefaultendpunct}{\mcitedefaultseppunct}\relax
\EndOfBibitem
\bibitem[Unke and Meuwly(2017)Unke, and Meuwly]{MM.rkhs:2017}
Unke,~O.~T.; Meuwly,~M. {Toolkit for the Construction of Reproducing
  Kernel-Based Representations of Data: Application to Multidimensional
  Potential Energy Surfaces}. \emph{J. Chem. Inf. Model} \textbf{2017},
  \emph{57}, 1923--1931\relax
\mciteBstWouldAddEndPuncttrue
\mciteSetBstMidEndSepPunct{\mcitedefaultmidpunct}
{\mcitedefaultendpunct}{\mcitedefaultseppunct}\relax
\EndOfBibitem
\bibitem[Ho and Rabitz(1996)Ho, and Rabitz]{ho96:2584}
Ho,~T.-S.; Rabitz,~H. {A general method for constructing multidimensional
  molecular potential energy surfaces from ab initio calculations}.
  \emph{J.~Chem.\ Phys.} \textbf{1996}, \emph{104}, 2584\relax
\mciteBstWouldAddEndPuncttrue
\mciteSetBstMidEndSepPunct{\mcitedefaultmidpunct}
{\mcitedefaultendpunct}{\mcitedefaultseppunct}\relax
\EndOfBibitem
\bibitem[Werner \latin{et~al.}(2019)Werner, Knowles, Knizia, Manby,
  {Sch\"{u}tz}, \latin{et~al.} others]{MOLPRO_brief}
Werner,~H.-J.; Knowles,~P.~J.; Knizia,~G.; Manby,~F.~R.; {Sch\"{u}tz},~M.,
  \latin{et~al.}  MOLPRO, version 2019.2, a package of ab initio programs.
  2019\relax
\mciteBstWouldAddEndPuncttrue
\mciteSetBstMidEndSepPunct{\mcitedefaultmidpunct}
{\mcitedefaultendpunct}{\mcitedefaultseppunct}\relax
\EndOfBibitem
\bibitem[Johnson(1978)]{johnson1978renormalized}
Johnson,~B.~R. The renormalized Numerov method applied to calculating bound
  states of the coupled-channel Schroedinger equation. \emph{J.~Chem.\ Phys.}
  \textbf{1978}, \emph{69}, 4678--4688\relax
\mciteBstWouldAddEndPuncttrue
\mciteSetBstMidEndSepPunct{\mcitedefaultmidpunct}
{\mcitedefaultendpunct}{\mcitedefaultseppunct}\relax
\EndOfBibitem
\bibitem[Gad\'ea \latin{et~al.}(1997)Gad\'ea, Berriche, Roncero, Villarreal,
  and Barrio]{gadea1997nonradiative}
Gad\'ea,~F.~X.; Berriche,~H.; Roncero,~O.; Villarreal,~P.; Barrio,~G.~D.
  Nonradiative lifetimes for LiH in the A state using adiabatic and diabatic
  schemes. \emph{J. Chem. Phys.} \textbf{1997}, \emph{107}, 10515--10522\relax
\mciteBstWouldAddEndPuncttrue
\mciteSetBstMidEndSepPunct{\mcitedefaultmidpunct}
{\mcitedefaultendpunct}{\mcitedefaultseppunct}\relax
\EndOfBibitem
\bibitem[Pawlak \latin{et~al.}(2017)Pawlak, Shagam, Klein, Narevicius, and
  Moiseyev]{pawlak2017adiabatic}
Pawlak,~M.; Shagam,~Y.; Klein,~A.; Narevicius,~E.; Moiseyev,~N. Adiabatic
  variational theory for cold atom--molecule collisions: Application to a
  metastable helium atom colliding with ortho-and para-hydrogen molecules.
  \emph{J.~Phys.\ Chem.~A} \textbf{2017}, \emph{121}, 2194--2198\relax
\mciteBstWouldAddEndPuncttrue
\mciteSetBstMidEndSepPunct{\mcitedefaultmidpunct}
{\mcitedefaultendpunct}{\mcitedefaultseppunct}\relax
\EndOfBibitem
\bibitem[Margulis \latin{et~al.}(2020)Margulis, Narevicius, and
  Narevicius]{margulis2020direct}
Margulis,~B.; Narevicius,~J.; Narevicius,~E. Direct observation of a Feshbach
  resonance by coincidence detection of ions and electrons in Penning
  ionization collisions. \emph{Nat. Comm.} \textbf{2020}, \emph{11}, 3553\relax
\mciteBstWouldAddEndPuncttrue
\mciteSetBstMidEndSepPunct{\mcitedefaultmidpunct}
{\mcitedefaultendpunct}{\mcitedefaultseppunct}\relax
\EndOfBibitem
\bibitem[Johnson()]{nlopt}
Johnson,~S.~G. The NLopt nonlinear-optimization package.
  \url{http://github.com/stevengj/nlopt}, Accessed: 2021-10-15\relax
\mciteBstWouldAddEndPuncttrue
\mciteSetBstMidEndSepPunct{\mcitedefaultmidpunct}
{\mcitedefaultendpunct}{\mcitedefaultseppunct}\relax
\EndOfBibitem
\bibitem[de~Fazio \latin{et~al.}(2012)de~Fazio, de~Castro-Vitores, Aguado,
  Aquilanti, and Cavalli]{faz12:244306}
de~Fazio,~D.; de~Castro-Vitores,~M.; Aguado,~A.; Aquilanti,~V.; Cavalli,~S.
  {The He + H$_2^+$ $\rightarrow$ HeH$^+$ + H reaction: Ab initio studies of
  the potential energy surface, benchmark time-independent quantum dynamics in
  an extended energy range and comparison with experiments}. \emph{J.~Chem.\
  Phys.} \textbf{2012}, \emph{137}, 244306\relax
\mciteBstWouldAddEndPuncttrue
\mciteSetBstMidEndSepPunct{\mcitedefaultmidpunct}
{\mcitedefaultendpunct}{\mcitedefaultseppunct}\relax
\EndOfBibitem
\bibitem[Aguado and Paniagua(1992)Aguado, and Paniagua]{agu92:1265}
Aguado,~A.; Paniagua,~M. {A New Functional form to Obtain Analytical Potentials
  of Triatomic Molecules}. \emph{J. Chem. Phys.} \textbf{1992}, \emph{96},
  1265--1275\relax
\mciteBstWouldAddEndPuncttrue
\mciteSetBstMidEndSepPunct{\mcitedefaultmidpunct}
{\mcitedefaultendpunct}{\mcitedefaultseppunct}\relax
\EndOfBibitem
\bibitem[Falcetta and Siska(1999)Falcetta, and Siska]{fal99:117}
Falcetta,~M.~F.; Siska,~P.~E. {The interaction between He and H$^{+}_{2}$:
  anisotropy, bond length dependence and hydrogen bonding}. \emph{Mol. Phys}
  \textbf{1999}, \emph{97}, 117--125\relax
\mciteBstWouldAddEndPuncttrue
\mciteSetBstMidEndSepPunct{\mcitedefaultmidpunct}
{\mcitedefaultendpunct}{\mcitedefaultseppunct}\relax
\EndOfBibitem
\bibitem[Bishop and Pipin(1995)Bishop, and Pipin]{bis95:15}
Bishop,~D.~M.; Pipin,~J. {Static electric properties of H and He}. \emph{Chem.\
  Phys.\ Lett.} \textbf{1995}, \emph{236}, 15\relax
\mciteBstWouldAddEndPuncttrue
\mciteSetBstMidEndSepPunct{\mcitedefaultmidpunct}
{\mcitedefaultendpunct}{\mcitedefaultseppunct}\relax
\EndOfBibitem
\bibitem[Velilla \latin{et~al.}(2008)Velilla, Lepetit, Aguado, Beswick, and
  Paniagua]{vel08:084307}
Velilla,~L.; Lepetit,~B.; Aguado,~A.; Beswick,~A.; Paniagua,~M. The H$_{3}^{+}$
  rovibrational spectrum revisited with a global electronic potential energy
  surface. \emph{J.~Chem.\ Phys.} \textbf{2008}, \emph{129}, 084307\relax
\mciteBstWouldAddEndPuncttrue
\mciteSetBstMidEndSepPunct{\mcitedefaultmidpunct}
{\mcitedefaultendpunct}{\mcitedefaultseppunct}\relax
\EndOfBibitem
\bibitem[Coxon and Hajigeorgiou(1999)Coxon, and Hajigeorgiou]{coxon:1999}
Coxon,~J.~A.; Hajigeorgiou,~P.~G. Experimental Born--Oppenheimer Potential for
  the X$^1 \Sigma^+$ Ground State of HeH$^+$: Comparison with the Ab Initio
  Potential. \emph{J.\ Mol.\ Struct.} \textbf{1999}, \emph{193}, 306--318\relax
\mciteBstWouldAddEndPuncttrue
\mciteSetBstMidEndSepPunct{\mcitedefaultmidpunct}
{\mcitedefaultendpunct}{\mcitedefaultseppunct}\relax
\EndOfBibitem
\bibitem[Dinelli \latin{et~al.}(1995)Dinelli, Le~Sueur, Tennyson, and
  Amos]{dinelli:1995}
Dinelli,~B.~M.; Le~Sueur,~C.~R.; Tennyson,~J.; Amos,~R.~D. Ab initio
  ro-vibrational levels of H$_3^+$ beyond the Born-Oppenheimer approximation.
  \emph{Chem.\ Phys.\ Lett.} \textbf{1995}, \emph{232}, 295--300\relax
\mciteBstWouldAddEndPuncttrue
\mciteSetBstMidEndSepPunct{\mcitedefaultmidpunct}
{\mcitedefaultendpunct}{\mcitedefaultseppunct}\relax
\EndOfBibitem
\bibitem[Balakrishnan \latin{et~al.}(1992)Balakrishnan, Smith, and
  Stoicheff]{balakrishnan:1992}
Balakrishnan,~A.; Smith,~V.; Stoicheff,~B. Dissociation energy of the hydrogen
  molecule. \emph{Phys.\ Rev.\ Lett.} \textbf{1992}, \emph{68}, 2149\relax
\mciteBstWouldAddEndPuncttrue
\mciteSetBstMidEndSepPunct{\mcitedefaultmidpunct}
{\mcitedefaultendpunct}{\mcitedefaultseppunct}\relax
\EndOfBibitem
\bibitem[Roy(2002)]{level}
Roy,~R.~L. LEVEL 7.5: a Computer Program to Solve the Radial Schr\"odinger
  Equation for Bound and Quasibound Levels. 2002\relax
\mciteBstWouldAddEndPuncttrue
\mciteSetBstMidEndSepPunct{\mcitedefaultmidpunct}
{\mcitedefaultendpunct}{\mcitedefaultseppunct}\relax
\EndOfBibitem
\end{mcitethebibliography}

\appendix

\renewcommand{\thetable}{S\arabic{table}}
\renewcommand{\thefigure}{S\arabic{figure}}
\renewcommand{\theequation}{S\arabic{equation}}
\renewcommand{\thesection}{S\arabic{section}}

\section{Two-body potential fitting}
In this work, the two-body interaction energy for a molecule AB was
expressed as\cite{MM.heh2:2019,faz12:244306,agu92:1265}
\begin{equation}
    \mathcal{V}^{(2)}_{AB}(R_{AB}) =
    \dfrac{c_{0}e^{-\alpha_{AB}R_{AB}}}{R_{AB}} + \sum_{i=1}^{M}
    C_{i}\rho_{AB}^{i} + \mathcal{V}_{long}(\tilde{r})
    \label{eq:general_2b}
\end{equation}
In equation \ref{eq:general_2b}, $C_{i}$ are linear coefficients with
$C_{0}>0$ to assure that the diatomic potential remains repulsive ($V_{\rm AB}(r_{\rm AB})\rightarrow \infty$) for $r_{\rm AB} \rightarrow 0$ and $\rho_{\rm AB}^{i}$ is defined
as:
\begin{equation}
    \rho_{AB}^{i} = R_{AB}e^{-\beta^{(2)}_{AB} R_{AB}}
    \label{eq:ro_ab}
\end{equation}
The long range part of equation \ref{eq:general_2b} is written
as\cite{fal99:117}:
\begin{equation}
    \mathcal{V}_{long}(\tilde{r}) =
    -\dfrac{\alpha_{d}q^{2}}{2\tilde{r}^{4}} -
    \dfrac{\alpha_{q}q^{2}}{2\tilde{r}^{6}} -
    \dfrac{\alpha_{o}q^{2}}{2\tilde{r}^{8}} -
    \dfrac{\beta_{ddq}q^{3}}{6\tilde{r}^{7}}-\dfrac{\gamma_{d}q^{4}}{24\tilde{r}^{8}}
    \label{eq:2b_longrange}
\end{equation}
In equation \ref{eq:2b_longrange}, $q$ is the charge, $\alpha_{i} ,\ i
\in \{d,q,o\}$ are the dipole, quadrupole and octupole
polarizabilities for H and He atoms, respectively, and $\beta_{ddq}$
and $\gamma_{d}$ are the first and second hyperpolarizabilities,
respectively. Values for these parameters were taken from
Refs. \citenum{fal99:117,bis95:15}. Finally, the coordinate
$\tilde{r}$, whose objective is to remove the divergence of the
long-range terms at short separations of H-H and H-He, is defined
as\cite{vel08:084307}
\begin{equation}
    \tilde{r} = r + r_{l}\exp{(-(r-r_{e}))}
\end{equation}
Here, $r_{l}$ is a distance parameter, and $r_{e}$ is the equilibrium
bond distance of the diatomic molecule. The linear coefficients
$C_{i}$ and the parameters $\alpha_{AB}$ and $\beta_{AB}^{(2)}$ in
equations \ref{eq:2b_longrange} and \ref{eq:ro_ab} were taken from
Ref. \citenum{MM.heh2:2019} for FCI and MRCI potentials. For the MP2
potential, the values were determined using the same method as
described in Ref. \citenum{MM.heh2:2019} and are given in Table
\ref{tab:2bodymp2coef}.
\begin{table}[h!]
\begin{tabular}{ccc}
\hline
Coefficients & H$_{2}^{+}$ & HeH$^{+}$   \\ \hline \hline
C$_{0}$      & 1.01921689  & 10.7720718  \\
$\alpha$     & 1.64361518  & 2.37373920  \\
C$_{1}$      & -0.73449753 & -4.46348296 \\
C$_{2}$      & 5.10324938  & 59.1487168  \\
C$_{3}$      & -81.5643228 & -3857.67751 \\
C$_{4}$      & 847.329344  & 104277.881  \\
C$_{5}$      & -5377.78872 & -1890643.12 \\
C$_{6}$      & 21730.0685  & 21015000.8  \\
C$_{7}$      & -56454.3034 & -1.3583E+08 \\
C$_{8}$      & 91470.4779  & 4.1364E+08  \\
C$_{9}$      & -84131.3637 & -29244093.5 \\
C$_{10}$     & 33516.3571  & -2.0736E+09 \\
$\beta$      & 0.99789130  & 2.23414441 \\ \hline \hline
\end{tabular}
\caption{Coefficients for the MP2/aug-cc-pV5Z diatomic
potentials.}
\label{tab:2bodymp2coef}
\end{table}

\newpage
\section{Discussion of morphing M1 for MRCI and MP2 PESs}

\noindent
{\it Multi-Reference CI:} Figure \ref{fig:spectra_all}B compares the
cross sections from experiments with the results from computations
with PESs before and after morphing M1 for the MRCI+Q PES. Overall,
the RMSE for the energies changes from 10.3 to 12.2 cm$^{-1}$, whereas
the intensities improve from an RMSE of 23.9 to 21.9 arb. u. The results
indicate that M1 has the most pronounced impact on intermediate values
of $j'$ (i.e. $j'=4,5$); see Figures \ref{fig:spectra_all}D and
E. Changes in the peak energies do not show a clear trend. The largest
improvements are observed for $j'=5$ and for $j'=[0,1]$. Errors for
peaks with $j'=8$ and $j'=6$ do not reduce using M1. The remaining
peaks showed an increase in the error after applying M1. For the peak
intensity, again, the largest improvement is observed for the
$j'=[0,1]$ peak. For most other peaks, with the exception of $j'=5$
and $j'=8$, there is clearly an improvement in the intensities.\\

\begin{figure}
    \centering
    \includegraphics[width=\textwidth]{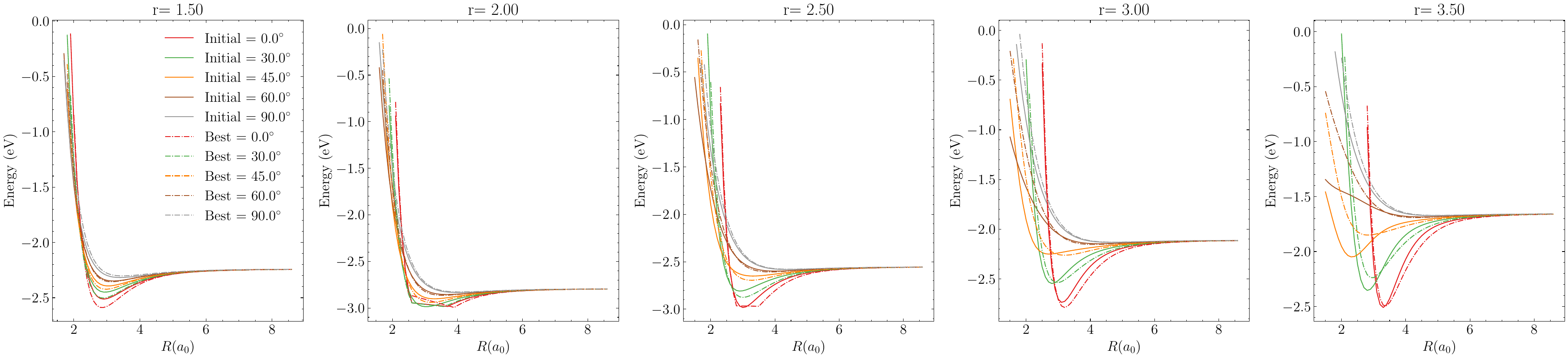}
    \caption{1D cuts of the MRCI PES obtained from the M1 procedure
      along $R$ for fixed $r_{\rm HH}$ and different angles ($\theta$).}
    \label{sifig:mrci.vdw.1d}
\end{figure}

\begin{figure}
    \centering \includegraphics[width=\textwidth]{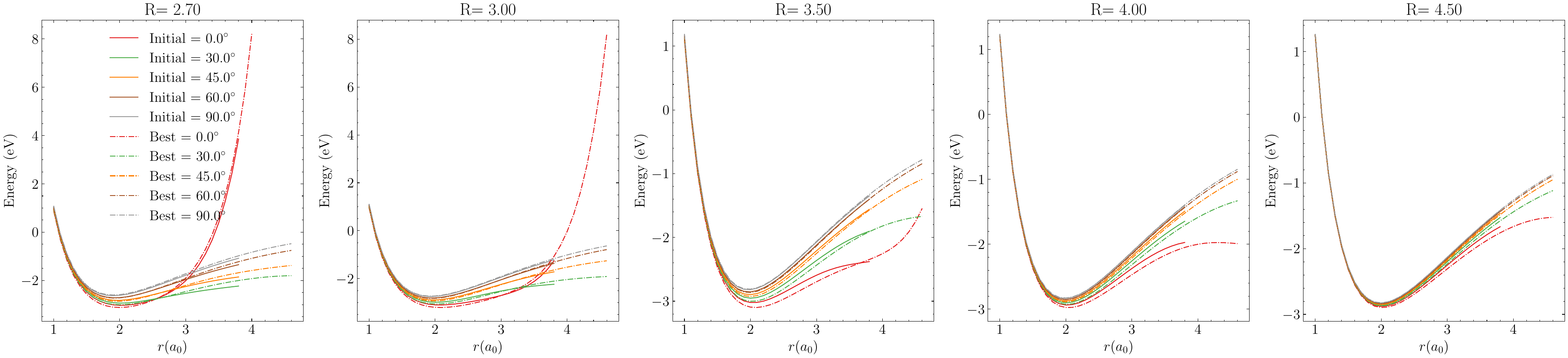}
    \caption{1D cuts of the MRCI PES obtained from the M1 procedure
      along $r_{\rm HH}$ for fixed $R$ and different angles ($\theta$).}
    \label{sifig:mrci.r.1d}
\end{figure}

\noindent
The initial and morphed MRCI PESs are compared in Figure
\ref{fig:surfaces_all}B. In this case, morphing increases the
anisotropy at long-range compared to the initial PES. However, changes
are more pronounced than for the FCI PES. One-dimensional cuts along
the $r_{\rm HH}$ and $R$ coordinates for given angle $\theta$ are provided in
Figures \ref{sifig:mrci.vdw.1d} and \ref{sifig:mrci.r.1d}. As for the
FCI PES, the difference between the initial surface and the morphed
surface are more pronounced as $r_{\rm HH}$ increases. The 1D cuts of the
surface at different values of $r_{\rm HH}$ (Figure \ref{sifig:mrci.r.1d}) show
further evidence of the change in the depth of the potential well. The
modifications of the energy curves with respect to the $r_{\rm HH}$ coordinate
follow the same trend as the FCI surface.\\

\noindent
{\it MP2:} The results for the lowest-quality surface (MP2) are shown
in Figures \ref{fig:spectra_all}C and \ref{fig:surfaces_all}C. The
RMSE for the energies improves from 13.1 to 12.8 cm$^{-1}$ whereas for
the intensities, it changes from 22.4 to 10.9 arb. u. Particularly notable is
the improvement in the intensities by more than a factor of
two. Overall, the changes in the position of the energies and the
intensities of the peaks for the calculated cross sections are more
pronounced than for the FCI and MRCI+Q PESs. The energy position for
peaks with large $j'$ ($j'=7$ and $j'=8$) improve by $\approx 5$
cm$^{-1}$. Another difference is that the shoulder of the peak at
$j'=8$ that appears for the two previously described surfaces is not
visible for the MP2 surface. For the peaks with $j'=4$ and $j'=5$, the
error with respect to the experimental spectra upon morphing increases
slightly.\\

\begin{figure}
    \centering \includegraphics[scale=0.6]{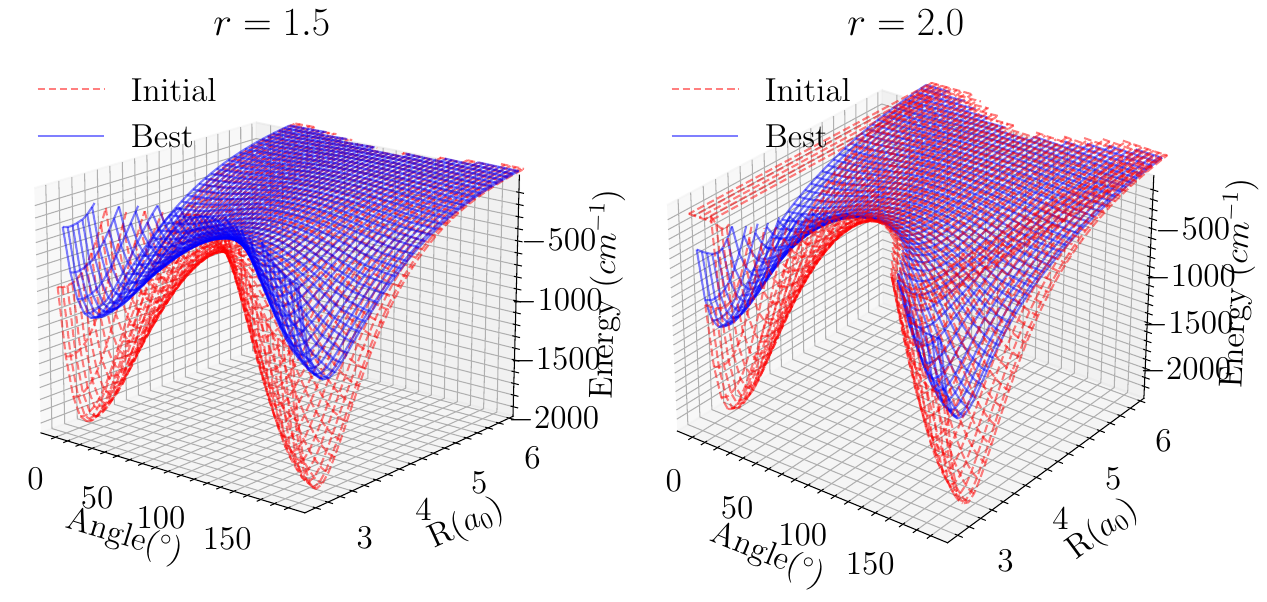}
    \caption{MP2 Potential energy surface from M1 procedure in 3D
      representation for fixed values of $r_{\rm HH}$, on the top of the figure
      is indicated the value of $r_{\rm HH}$. As zero of energy, it was
      considered the value at the given $r_{\rm HH}$ and $R=\infty$. Energies
      are in cm$^{-1}$ and separated by 100 cm$^{-1}$}
    \label{sifig:surface_mp2_3d}
\end{figure}

\begin{figure}
    \centering \includegraphics[width=\textwidth]{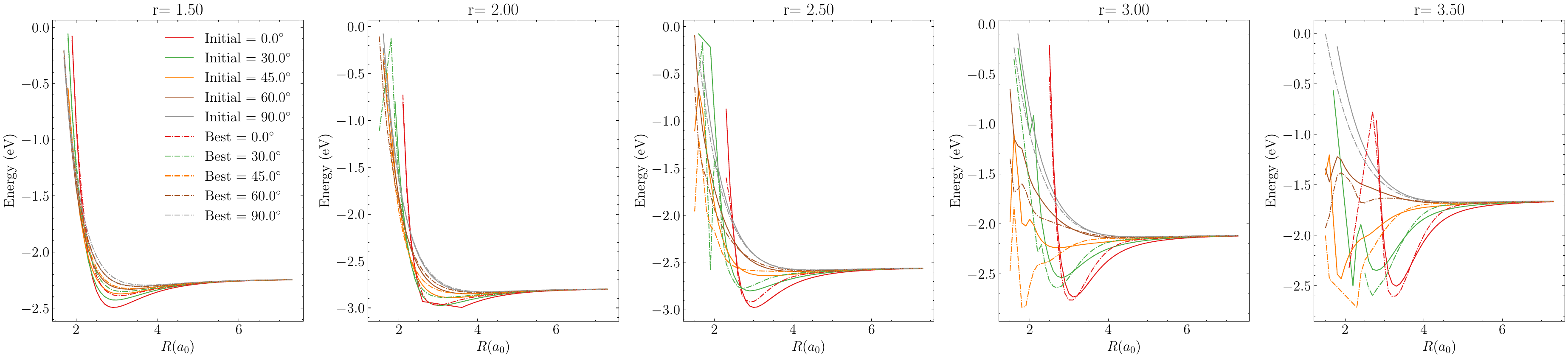}
    \caption{1D cuts of the MP2 PES obtained from the M1 procedure
      along $R$ for fixed $r_{\rm HH}$ and different angles ($\theta$).}
    \label{sifig:mp2.vdw.1d}
\end{figure}

\begin{figure}
    \centering
    \includegraphics[width=\textwidth]{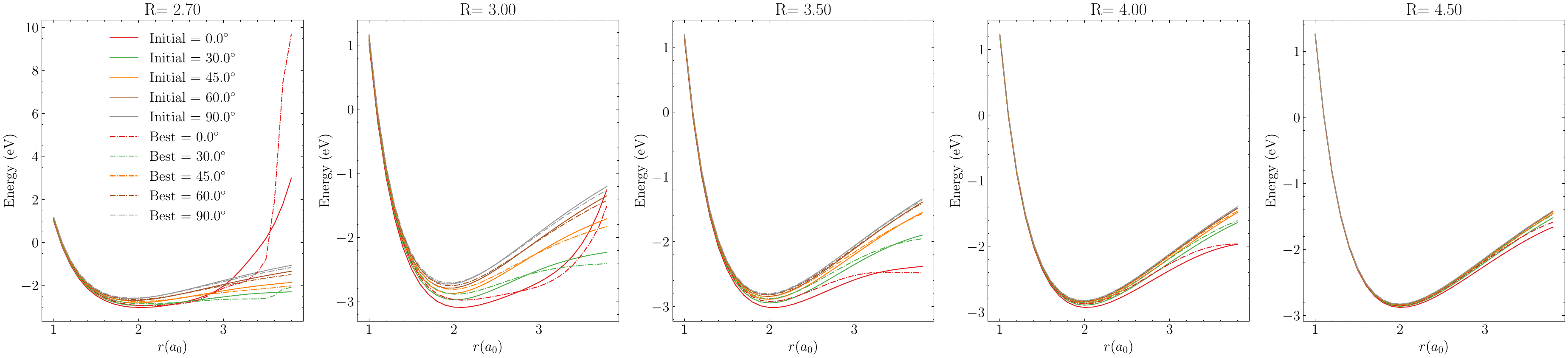}
    \caption{1D cuts of the MP2 PES obtained from the M1 procedure
      along $r_{\rm HH}$ for fixed $R$ and different angles ($\theta$).}
    \label{sifig:mp2.r.1d}
\end{figure}

\begin{figure}
    \centering
    \includegraphics[width=\textwidth]{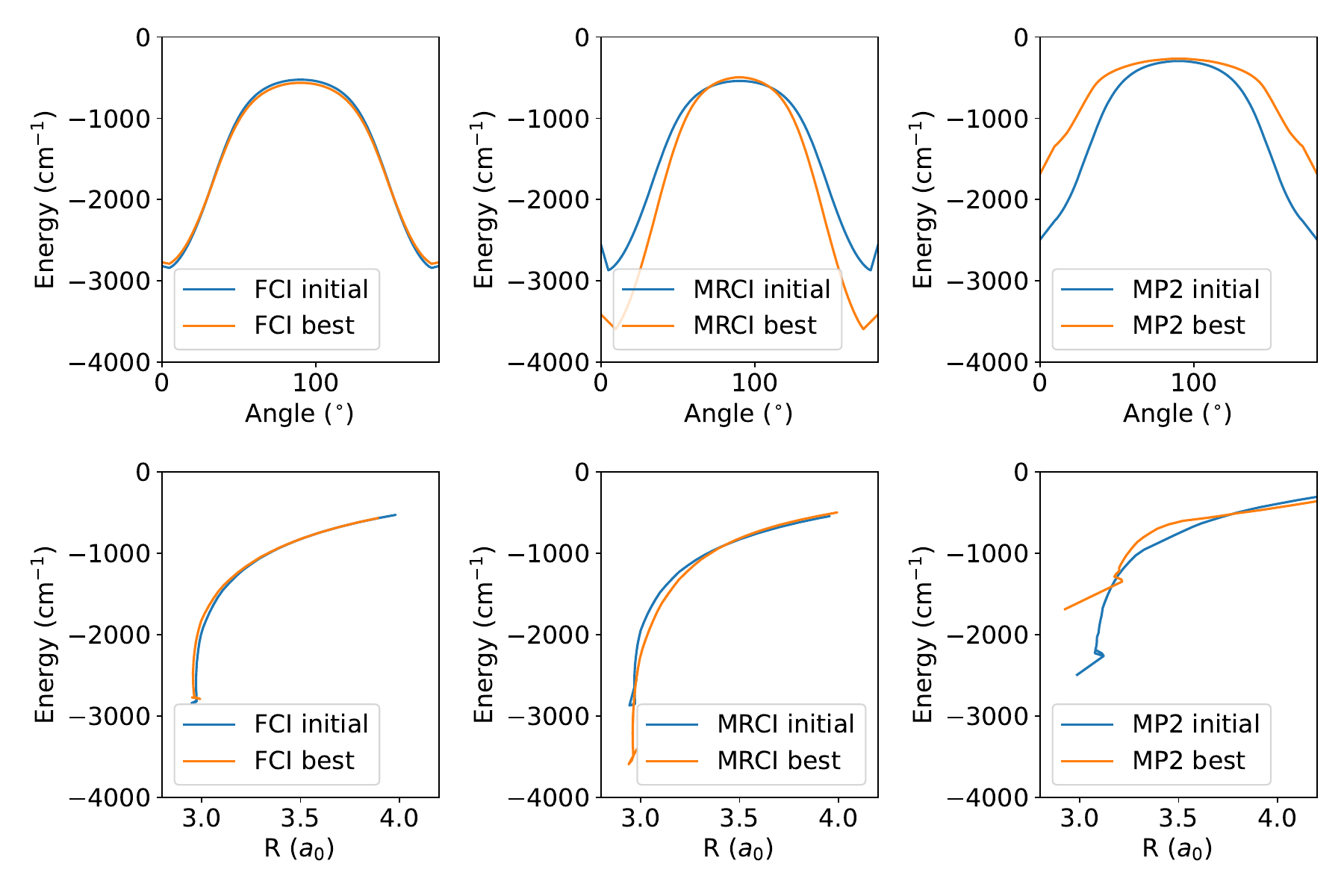}
    \caption{Minimum energy paths (MEPs) for the different surfaces
      obtained from the M1 method with respect to the variables
      $\theta$ and $R$. The zero of energy of the path is the energy of the separated monomers.}
    \label{sifig:meps_m1}
\end{figure}

\begin{figure}
    \centering
    \includegraphics[width=\textwidth]{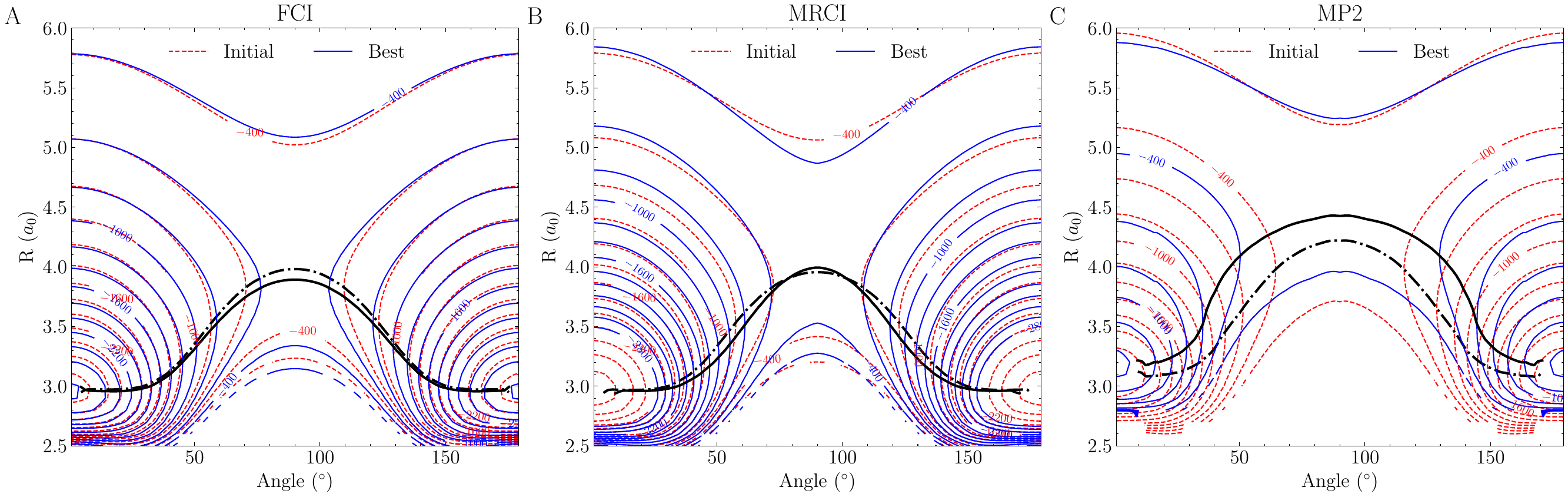}
    \caption{Minimum energy paths (MEPs) for the different surfaces
      studied before and after applying the morphing M1 method. In
      solid black, it is shown the MEP for the morph PES. In dotted
      black, the MEP is shown for the initial PES.}
    \label{sifig:meps_in_pes_m1}
\end{figure}

\noindent
The original MP2 PES and its morphed variant for a H$_2^+$ separation
of $r_{\rm HH}=2.0$ a$_{0}$ are reported in Figure
\ref{fig:surfaces_all}C. Because M{\o}ller-Plesset second-order theory
is a single-reference method and makes further approximations, the
changes in the topology of the PES are considerably larger than for
the FCI and MRCI+Q PESs. Most of the isocontours are compressed
compared with the initial MP2 surface, and the well depth is reduced
from 2493 cm$^{-1}$ to 1684 cm$^{-1}$ (Table \ref{tab:deeps}), see
Figure \ref{sifig:surface_mp2_3d}. The one-dimensional cuts along the
$r_{\rm HH}$ and $R$ coordinates for given $\theta$, see Figures
\ref{sifig:mp2.vdw.1d} and \ref{sifig:mp2.r.1d}, show that as $r_{\rm HH}$
increases the single-reference assumption of the method, leading to
convergence problems for small $R$. As a consequence of the
contraction of the potential wells, the barrier of the transition
state at $\theta \approx 90^\circ$ is increased, which is further
confirmed by the Minimum Energy Path (MEP) shown in Figure
\ref{sifig:meps_m1}C. A more detailed analysis of the MEP (Figure
\ref{sifig:meps_in_pes_m1}C) reveals a small increase in the energy of the
transition state along the angular coordinate $\theta$. On the other
hand, for the $R-$coordinate a non-physical barrier emerges at around
3.5 $a_{0}$.  \\

\newpage
\section{Resonances under Morphing}
The cross sections depending on the binding energy between He and
H$_2^+$ as opposed to the relative kinetic energy of the two
reactants shows distinct peaks that are no longer separated by final
states ($j'$) of the H$_2^+$ fragment but rather appear as one or
several Feshbach Resonances per input $J$ and $\ell$ at certain values
of the binding energy. Both the energy at which a Feshbach Resonance
appears, and the distribution of intensities in all exit channels,
depend sensitively on the topography of the PES. In consequence, the
effect of morphing on the PES can influence the number, energy and
intensities of the Feshbach resonances. To illustrate this, it is
instructive to consider projections of wave functions for particular
resonances to characterize how changes in the PES, which lead to
changes in the collision cross-section, are reflected in the radial
and angular behaviour of the wave function.\\

\begin{figure}
    \centering
    \begin{tikzpicture}
    \node [] (para) at (0,0)
          {\includegraphics[width=14cm,keepaspectratio]{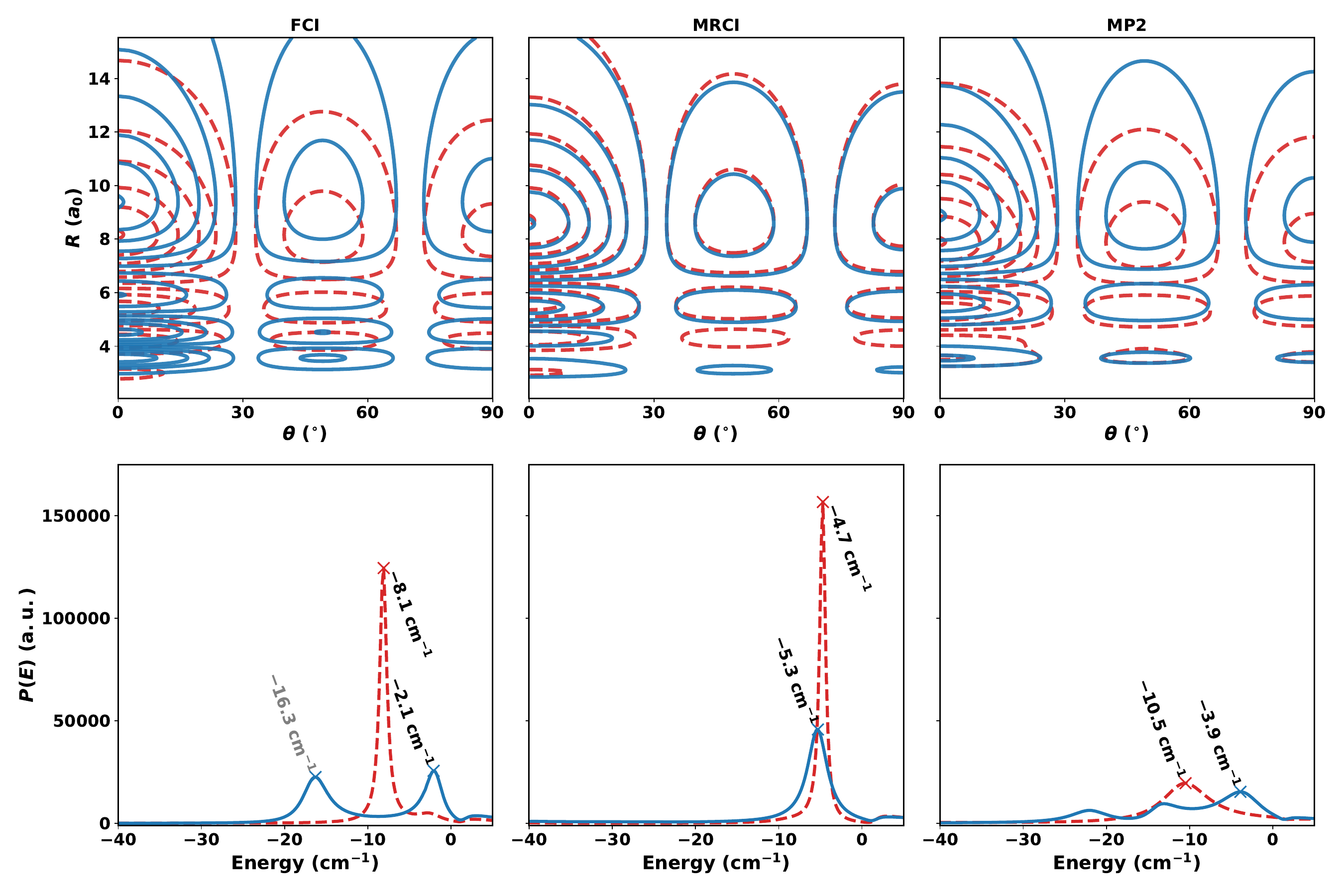}};

    \node [anchor=north east] (ortho) at (para.south east) {
      \includegraphics[width=14cm,keepaspectratio]{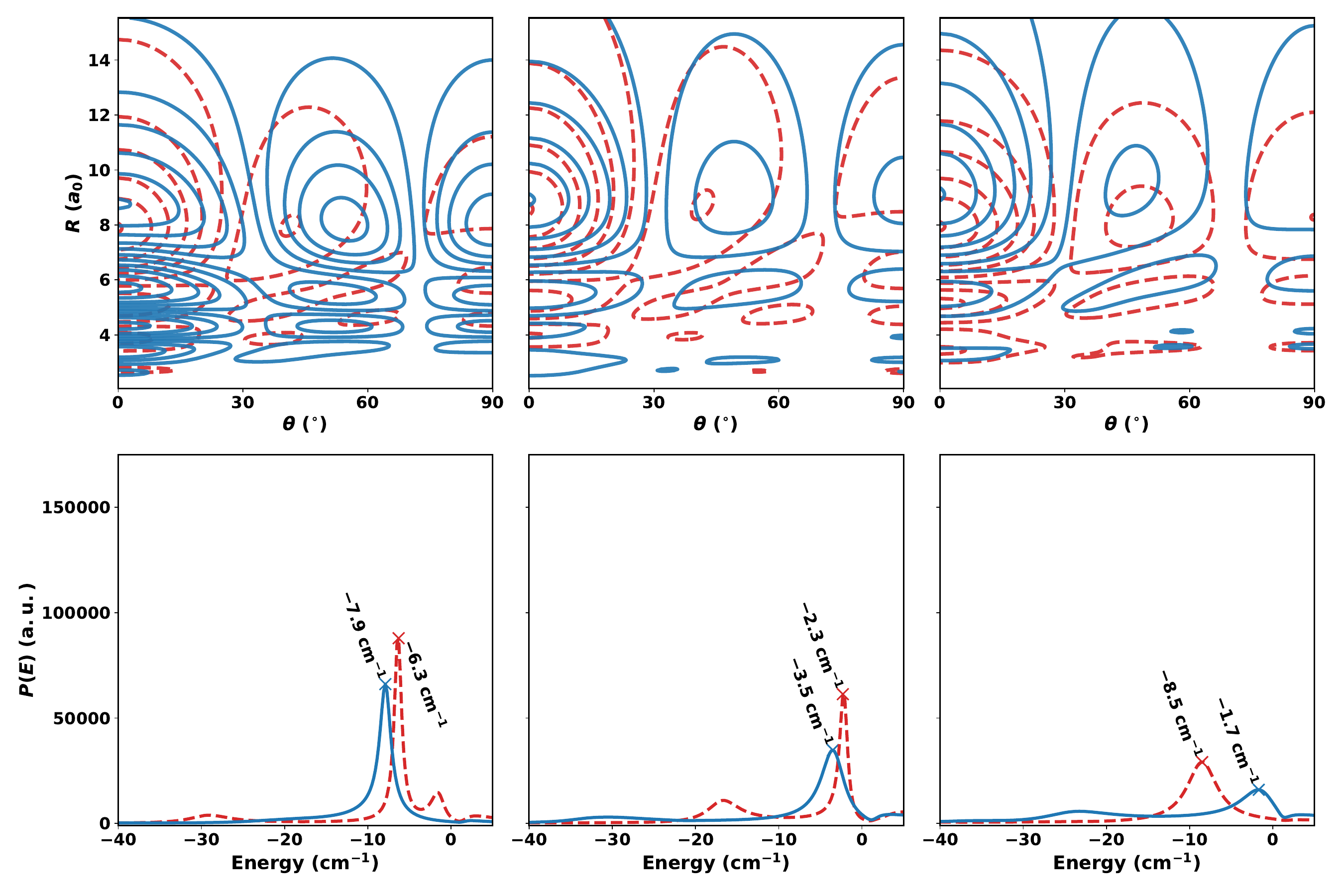}};
    \end{tikzpicture}
    \caption{Comparison of the unmorphed (red, dotted) and morphed
      (blue) absolute value squared resonance wave functions in two
      dimensions $(R,\theta)$ in the case of para-H$_2^+$ $\ell=4,
      J=4$ (upper two rows) and ortho-H$_2^+$ $\ell=4, J=5$ (lower two
      rows) for resonance energies as marked and labelled in the
      corresponding cross sections are shown as a function of binding
      energy (second and fourth rows for para and ortho,
      respectively).  The resonance wave functions have been scaled to
      have a maximal value of one, and the contours occur at 0.01,
      0.1, 0.25, 0.5, 0.75 and 0.99.}
    \label{sifig:wfs}
\end{figure}

\noindent
Figure \ref{sifig:wfs} shows the square of the $(v'=v)$ and $(j'=j)$
components of the resonance wave functions (first and third rows of
panels) and corresponding resonances in the cross-section (second and
fourth rows of panels) for the dominant $\ell$ and $J$ contributions
for para- and ortho-H$_2^+$ for all three unmorphed and morphed PESs,
respectively. The number, position(s) and intensities of the
spectroscopic features respond to morphing in a largely unpredictable
way. As an example, the unmorphed and morphed PESs at the FCI level
are considered for para-H$_2^+$ with $(\ell=4, J=4)$ (left column,
rows 1 and 2 in Figure \ref{sifig:wfs}). Although M1 changes the
topology of the morphed PES only in a minor fashion, the effect on the
wavefunctions and resulting spectroscopic feature is clearly
visible. For the unmorphed FCI PES there is one resonance at --8.1
cm$^{-1}$ which splits into two resonances at --2.1 cm$^{-1}$ and
--16.3 cm$^{-1}$ of approximately equal height upon morphing the
PES. Accordingly, the wavefunctions also differ, in particular in the
long-range part, i.e. for large $R$. Similar observations were made
for the wavefunctions on the MP2 PES, whereas for the MRCI PESs the
changes in the wavefunctions are comparatively smaller.\\

\noindent
Conversely, for ortho-H$_2^+$ the resonances of both FCI and MRCI PESs
are affected in a comparable fashion and more noticeable changes to
the resonance wave function are observed than for para-H$_2^+$. Whilst
the resonance wave functions are shifted to larger $R$ in the cases of
FCI and MP2, the MRCI resonance wave function only experiences a small
shift. Significantly, even though the anisotropy of the PESs only
changes in a minor fashion under morphing, all three resonance wave
functions respond owing to a change in the superposition of outgoing
partial wave (quantum number $\ell'$). For the FCI and MP2 PESs
angular/radial coupling is enhanced by morphing, which leads to
elongation of certain lobes in the wavefunctions along the
$(R,\theta)-$direction for ortho-H$_2^+$--He. This contrasts with
para-H$_2^+$--He for which unique assignments of the ro-vibrational
quantum numbers is possible from conventional node-counting.\\

\clearpage
\section{Figures}

\begin{figure}[h!]
    \centering \includegraphics[scale=0.6]{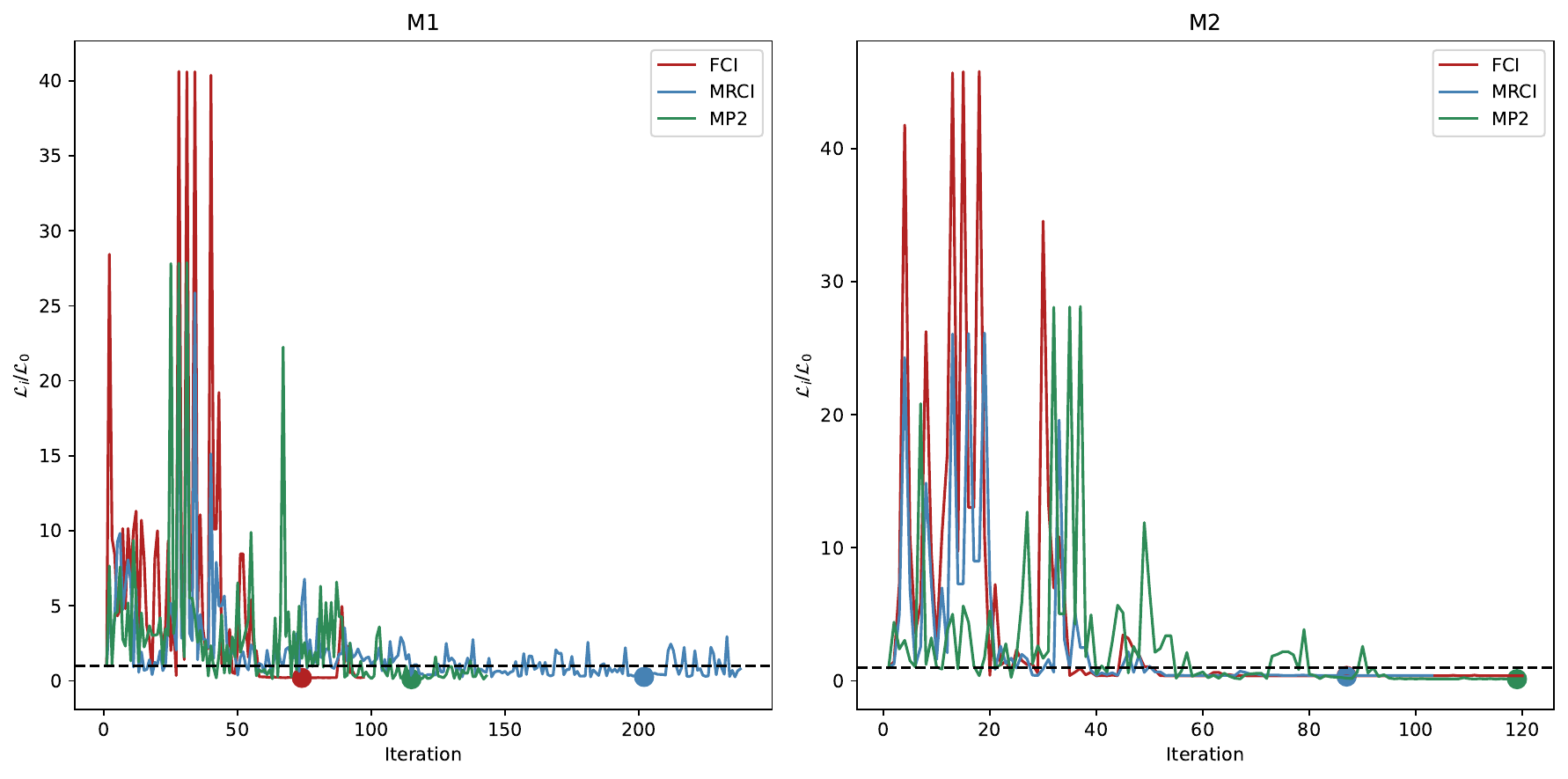}
    \caption{Evolution of the values of the loss function
      ($\mathcal{L}_{i}$) over the iterations with respect to the
      initial value ($\mathcal{L}_{0}$) for the M1(left) and M2 (right) methods.  The loss
      function is defined in the main manuscript.}
    \label{sifig:error_curve}
\end{figure}

\begin{figure}
    \centering \includegraphics[width=\textwidth]{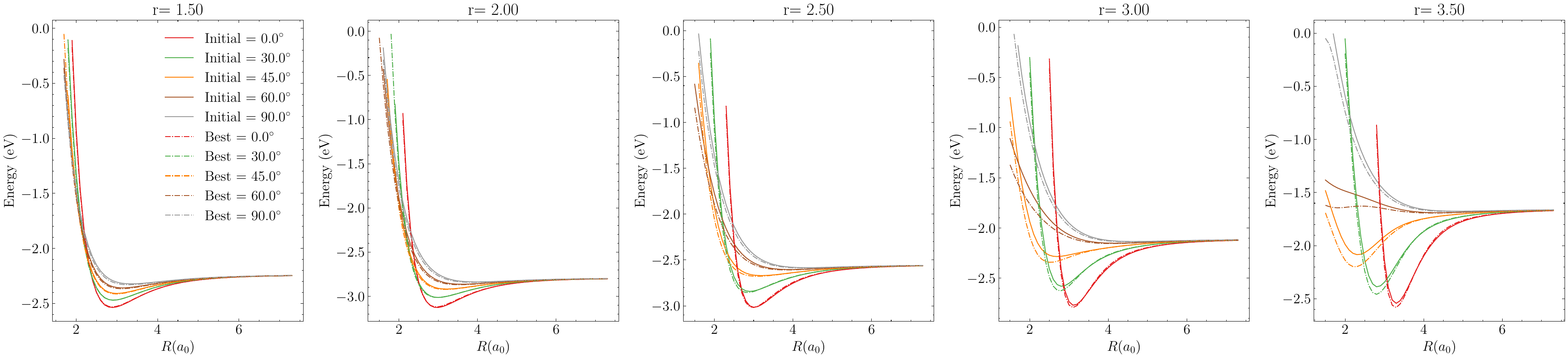}
    \caption{1D cuts of the FCI PES obtained from the M1 procedure
      along $R$ for fixed $r_{\rm HH}$ and different angles ($\theta$).}
    \label{sifig:fci.vdw.1d}
\end{figure}

\begin{figure}
    \centering
    \includegraphics[width=\textwidth]{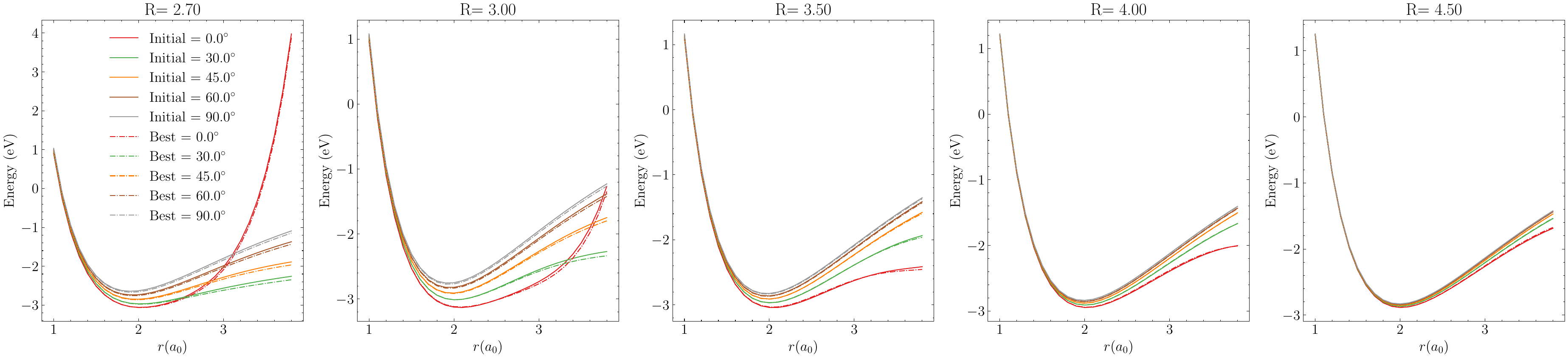}
    \caption{1D cuts of the FCI PES obtained from the M1 procedure
      along $r_{\rm HH}$ for fixed $R$ and different angles ($\theta$).}
    \label{sifig:fci.r.1d}
\end{figure}

\begin{figure}
    \centering \includegraphics[width=\textwidth]{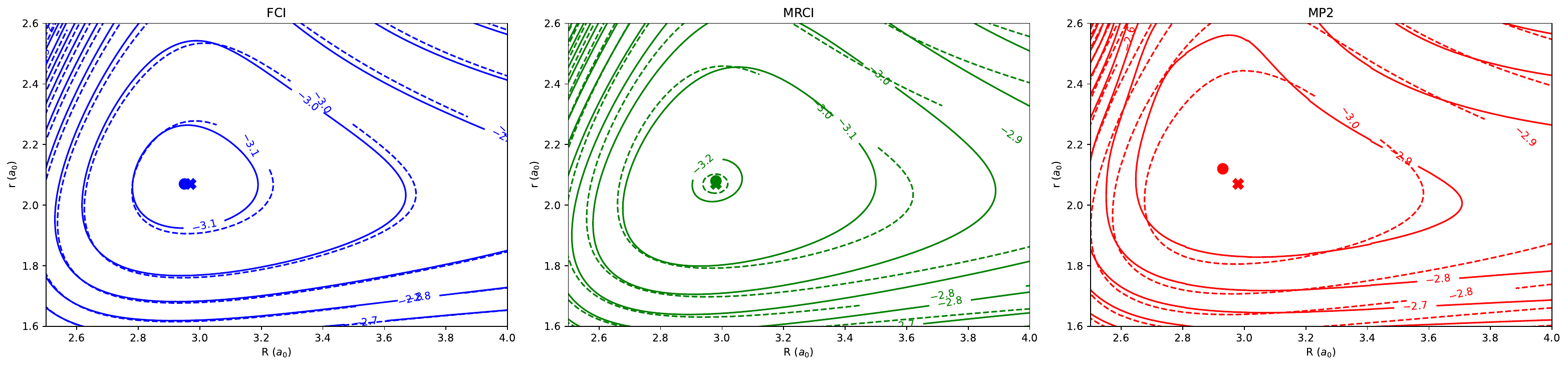}
    \caption{Potential energy surface 2D projection $V(R,r)$ at
      $\theta = 0^{\circ}$ obtained from the M1 procedure for the
      three potentials studied in this work. The dotted lines represent the unmorphed potential, complementary full lines show the morphed potential. Isocontours are separated by 0.1 eV. The minimum of the potential is indicated with a dot and a cross for the unmorphed and morphed potential.}
    \label{sifig:Rr_theta0}
\end{figure}

\begin{figure}
    \centering \includegraphics[width=\textwidth]{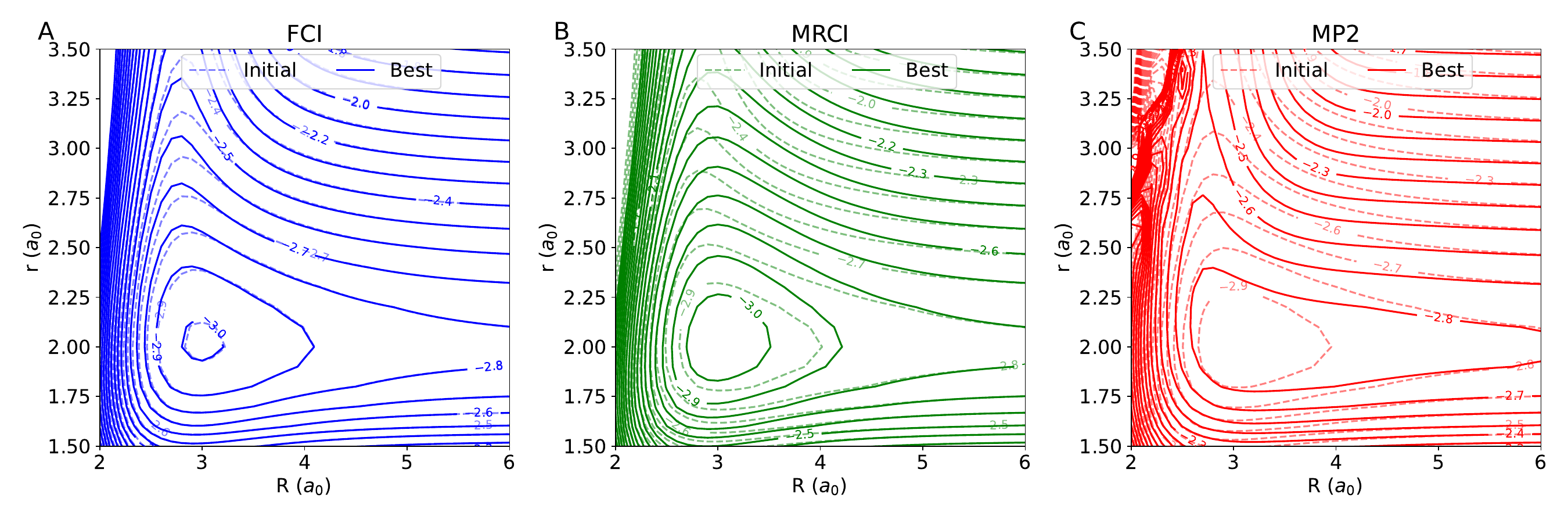}
    \caption{Potential energy surface 2D projection $V(R,r)$ at
      $\theta = 30^{\circ}$ obtained from the M1 procedure for the
      three potentials studied in this work. The dotted lines represent the unmorphed potential; complementary full lines show the morphed potential. Isocontours are separated by 0.1 eV. }
    \label{sifig:Rr_theta30}
\end{figure}

\begin{figure}
    \centering
    \includegraphics[width=\textwidth]{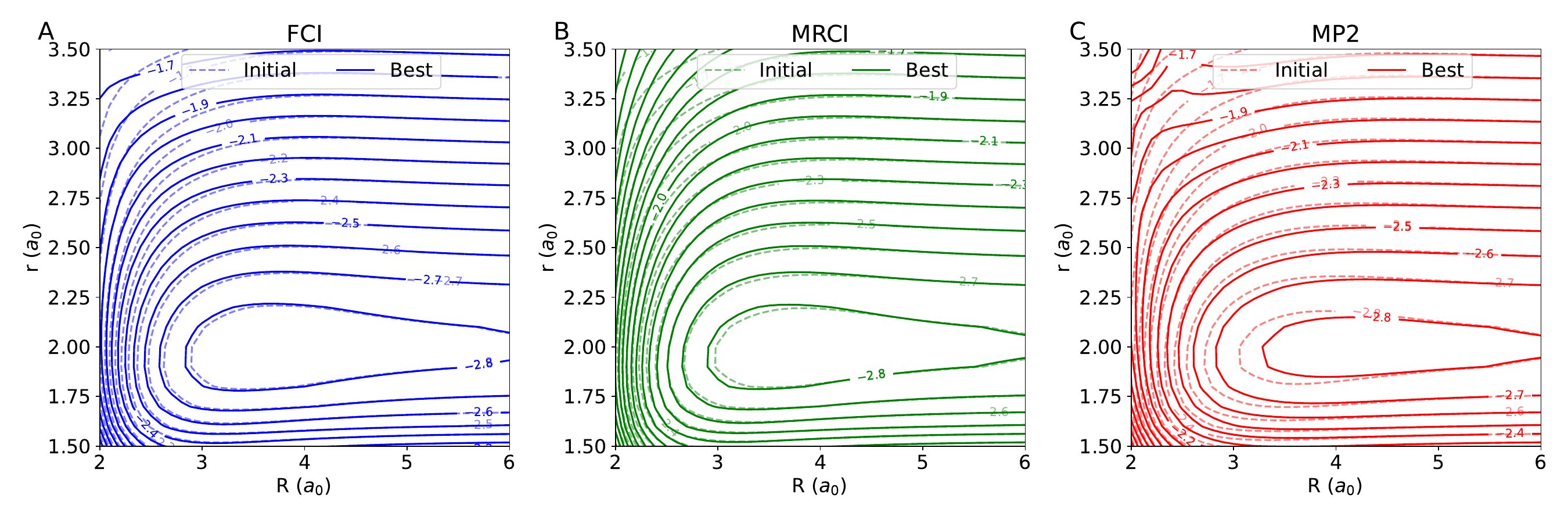}
    \caption{Potential energy surface 2D projection $V(R,r)$ at
      $\theta = 60^{\circ}$ obtained from the M1 procedure for the
      three potentials studied in this work. The dotted lines represent the unmorphed potential; complementary full lines show the morphed potential. Isocontours are separated by 0.1 eV.}
    \label{sifig:Rr_theta60}
\end{figure}
\pagebreak
\begin{figure}
    \centering
    \includegraphics[scale=0.7]{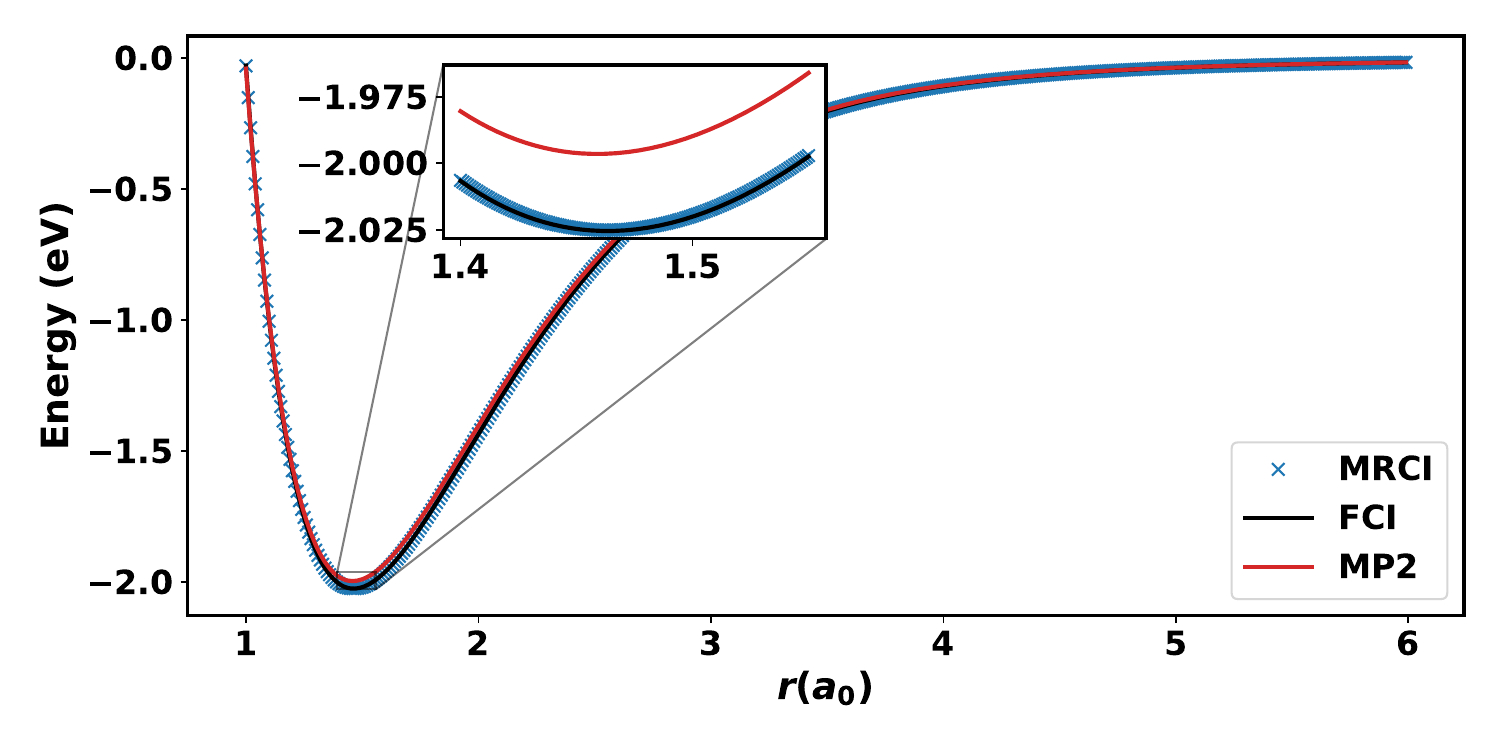}
    \caption{2-body potential for the HeH$^{+}$ molecule. The inset
      shows the region close to the equilibrium geometry. Zero of
      energy is at the He+H$^+$ asymptote. For HeH$^+$ the
      experimentally determined\cite{coxon:1999} dissociation energy
      is $D_e = 16456.23$ cm$^{-1}$ compares with 16406.04 (FCI),
      16403.96 (MRCI+Q), and 16171.65 (MP2) cm$^{-1}$ from the fitted
      2-body potentials. The difference of 235 cm$^{-1}$ between FCI
      and MP2 is substantial. Remaining differences between experiment
      and FCI are due to basis set incompleteness and
      non-Born-Oppenheimer effects, not included in the
      calculations. For other systems, such effects have been
      estimated at several ten up to 100 cm$^{-1}$ on total
      energies.\cite{dinelli:1995}}
    \label{sifig:2body_hehp}
\end{figure}

\begin{figure}
    \centering
    \includegraphics[scale=0.7]{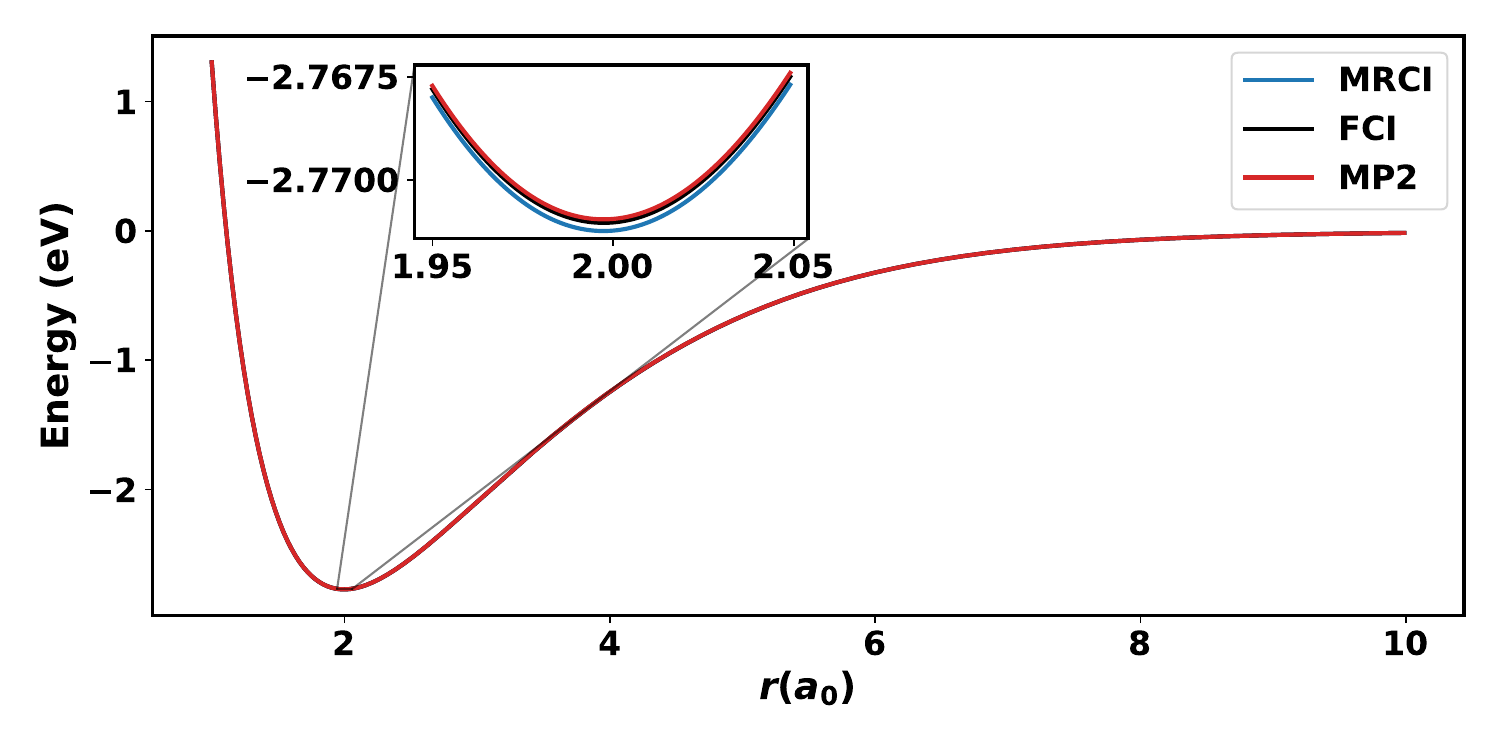}
    \caption{2-body potential for the H$_{2}^{+}$ molecule.  The zero
      of energy is the H+H$^+$ asymptote and the inset shows the
      region close to the equilibrium geometry. The experimentally
      determined\cite{balakrishnan:1992} dissociation energy $D_0 =
      21379.36 \pm 0.08$ cm$^{-1}$ compares with 21428.5 (FCI),
      21430.1 (MRCI+Q), and 21427.5 (MP2) cm$^{-1}$. The location of
      the ground states $(v=0, j=0)$ was determined using the LEVEL
      code.\cite{level} Remaining differences between experiment and
      FCI are due to basis set incompleteness and non-Born-Oppenheimer
      effects, not included in the calculations. For other systems,
      such effects have been estimated at several ten up to 100
      cm$^{-1}$ on total energies.\cite{dinelli:1995}}
    \label{sifig:2body_h2p}
\end{figure}

\begin{figure}
    \centering
    \includegraphics[width=\textwidth]{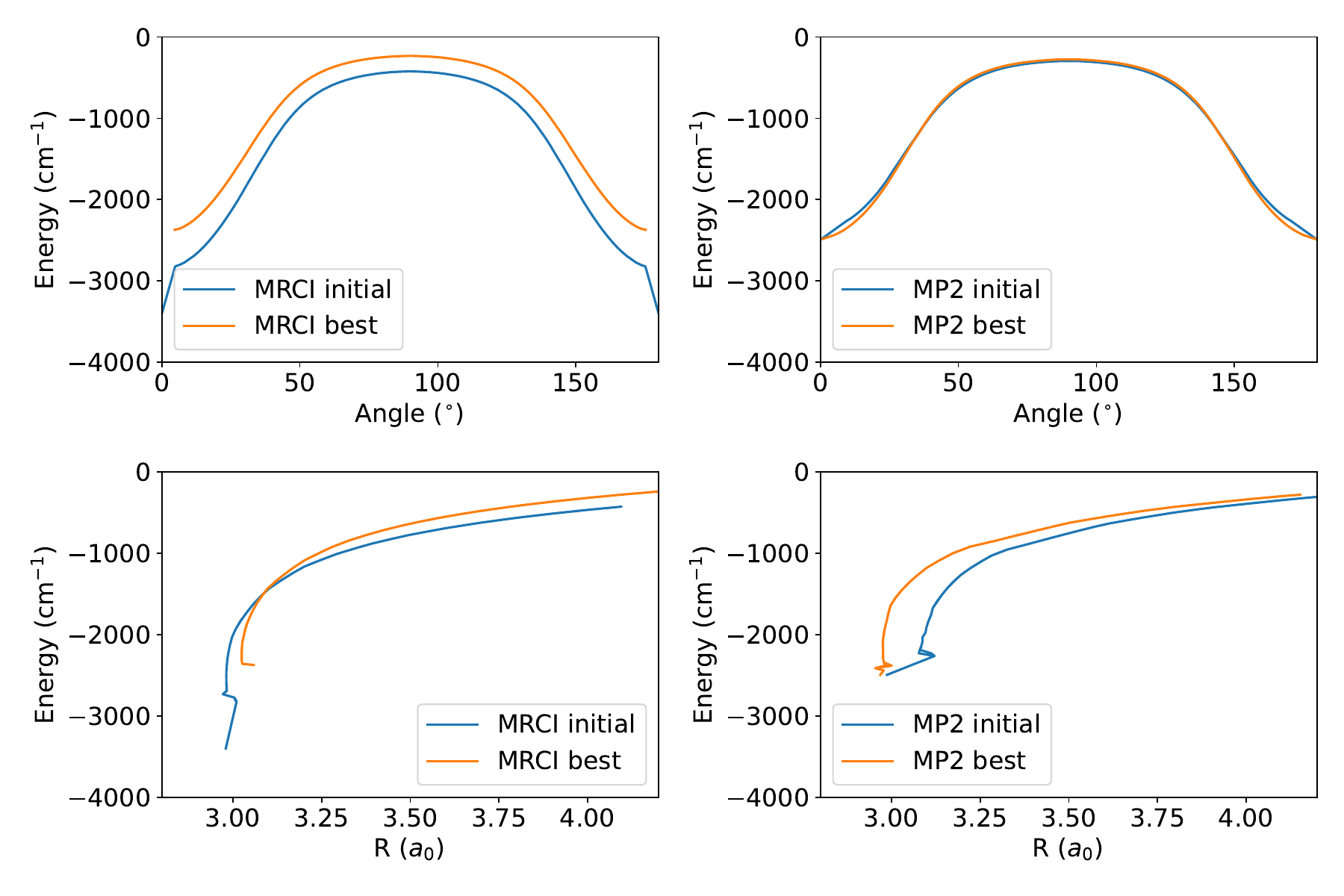}
    \caption{Minimum energy paths (MEPs) for the different surfaces
      obtained from the M2 method with respect to the variables
      $\theta$ and $R$. The zero of energy of the path is the energy of the separated monomers.}
    \label{sifig:meps_m2}
\end{figure}

\begin{figure}
    \centering
    \includegraphics[width=\textwidth]{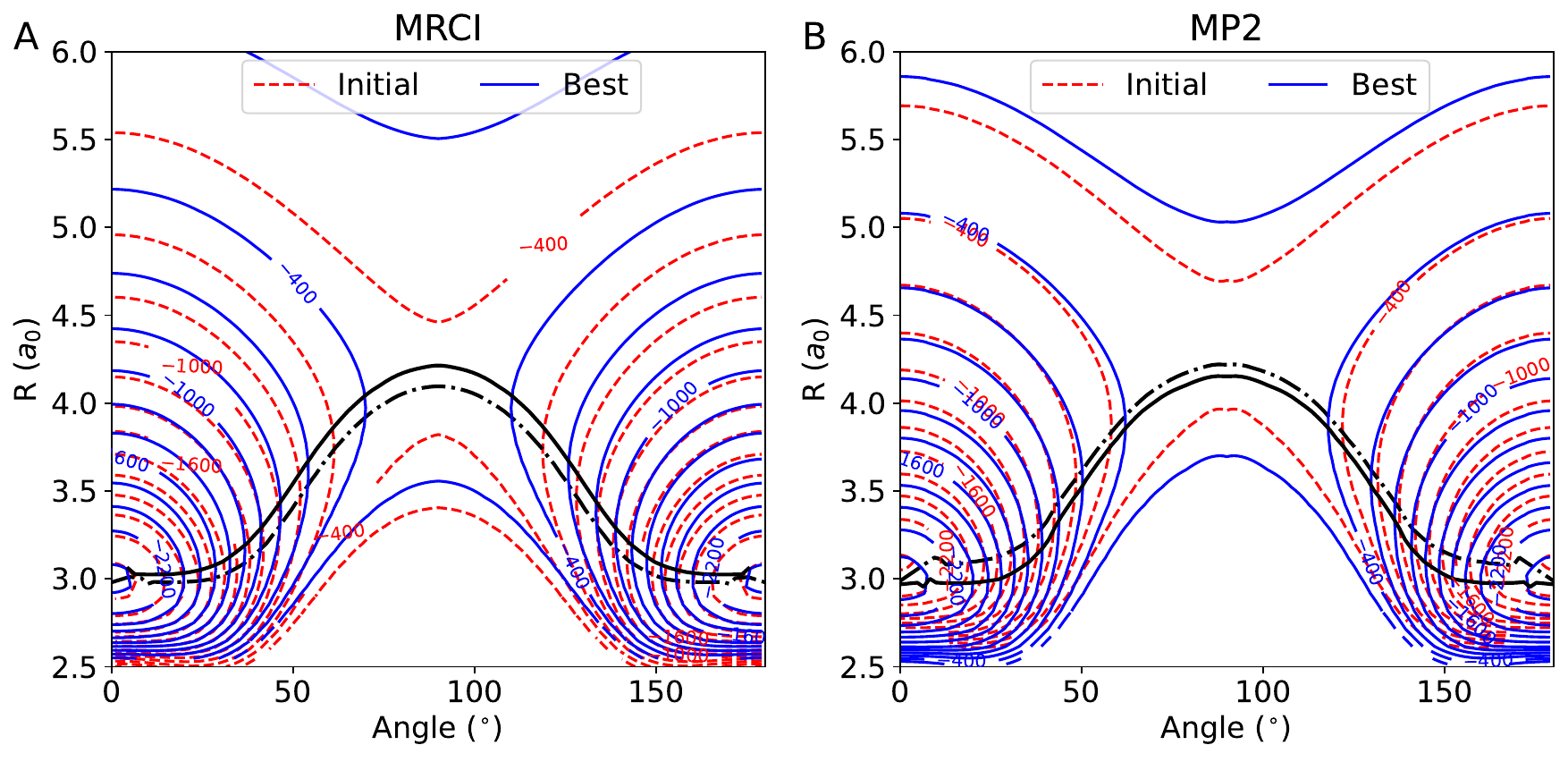}
    \caption{Minimum energy paths (MEPs) for the different surfaces
      studied before and after applying the morphing M2 method. In
      solid black, it is shown the MEP for the morph PES. In dotted
      black, the MEP is shown for the initial PES.}
    \label{sifig:meps_in_pes_m2}
\end{figure}

\begin{figure}
    \centering \includegraphics[width=\textwidth]{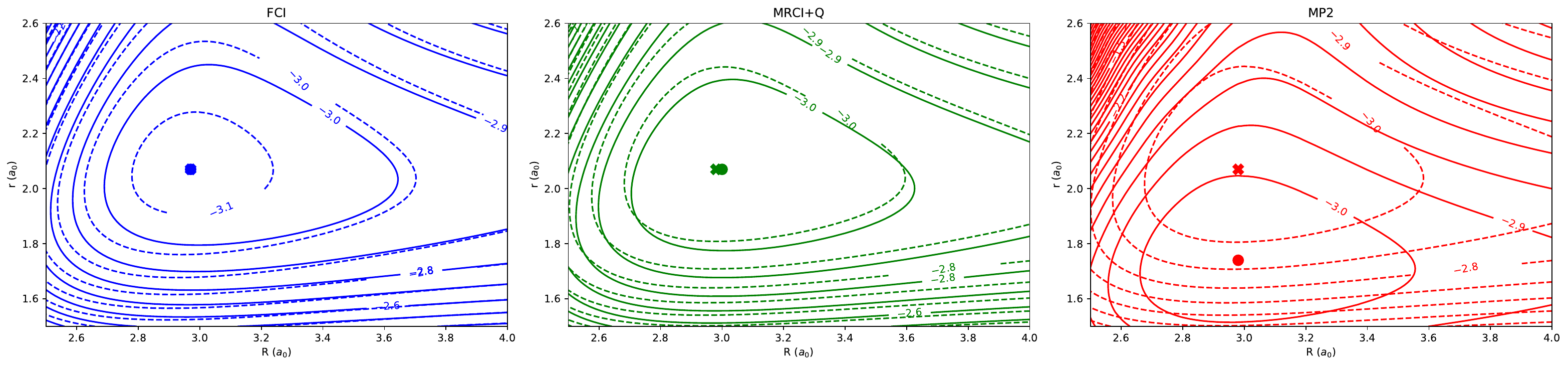}
    \caption{Potential energy surface 2D projection $V(R,r)$ at
      $\theta = 0^{\circ}$ obtained from the M2 procedure for the
      three potentials studied in this work. The dotted lines represent the unmorphed potential. Complementary full lines show the morphed potential. Isocontours are separated by 0.1 eV. The minimum of the potential is indicated with a dot and a cross for the unmorphed and morphed potential.}
    \label{sifig:Rr_M2}
\end{figure}

\begin{figure}
    \centering
    \includegraphics[width=\textwidth]{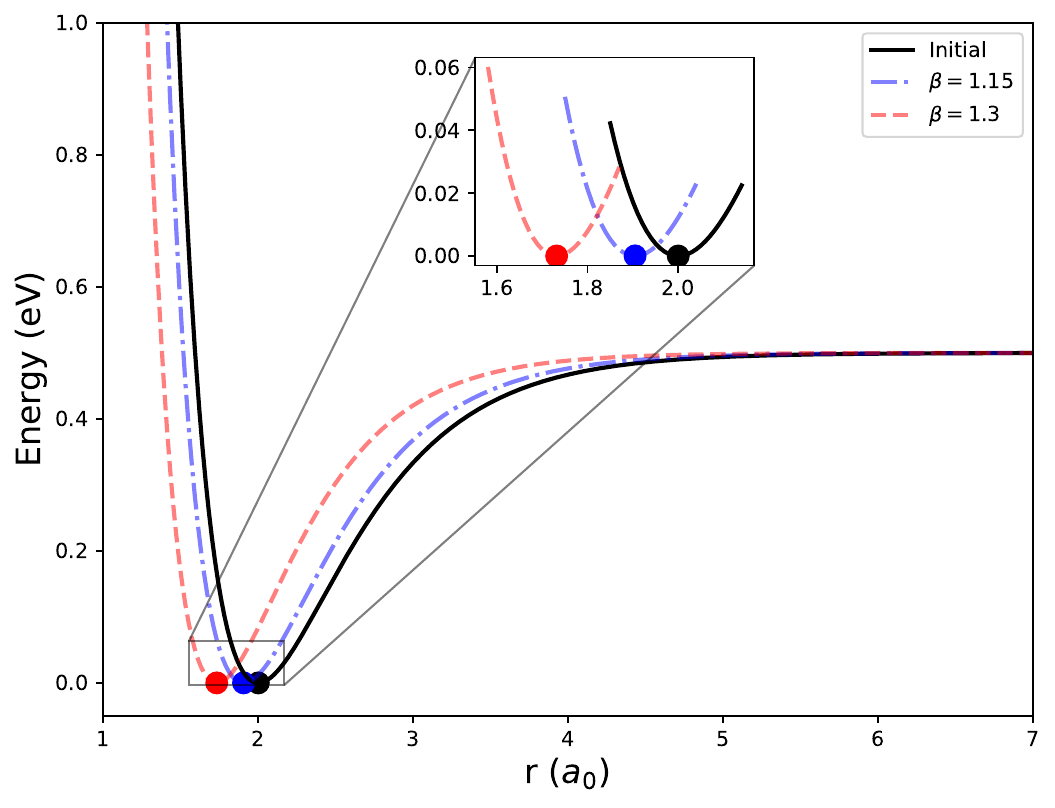}
    \caption{Schematic representation of the changes in the geometry by the morphing transformation for an example Morse potential curve. If the value of the scaling parameter $\beta>1$, the equilibrium minima will be displaced to smaller values of coordinate $r$. The insight shows the change in the curvature close to the value of the equilibrium geometry}
    \label{sifig:scheme}
\end{figure}

\begin{figure}
    \centering
    \includegraphics[scale=1]{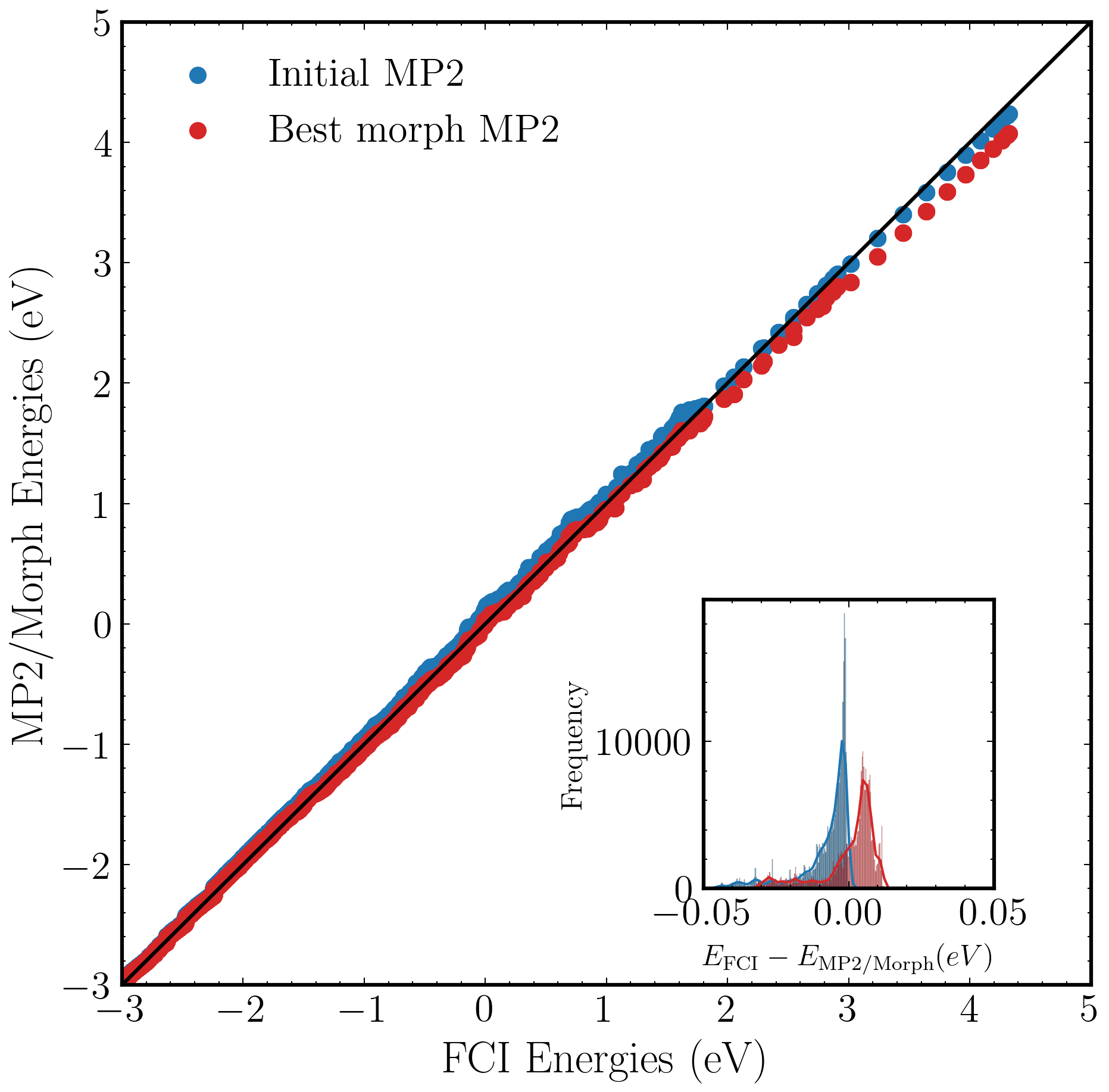}
    \caption{Scatter plot of the energies of the FCI surface vs the
      initial and morphed MP2 surface for the PES-to-PES procedure.
      In the inset of the figure, the distribution of differences
      $E_{\rm FCI}-E_{\rm MP2}$ for unmorphed and morphed MP2
      surfaces, respectively. The Pearson correlation coefficients are
      0.9984 and 0.9988 for unmorphed and morphed MP2 PESs and the
      RMSE decreases from 138 cm$^{-1}$ to 87 cm$^{-1}$ upon morphing
      for energies spanning $\sim 7$ eV.}
    \label{sifig:scatter_pes2pes}
\end{figure}

\begin{figure}
    \centering
    \includegraphics[width=\textwidth]{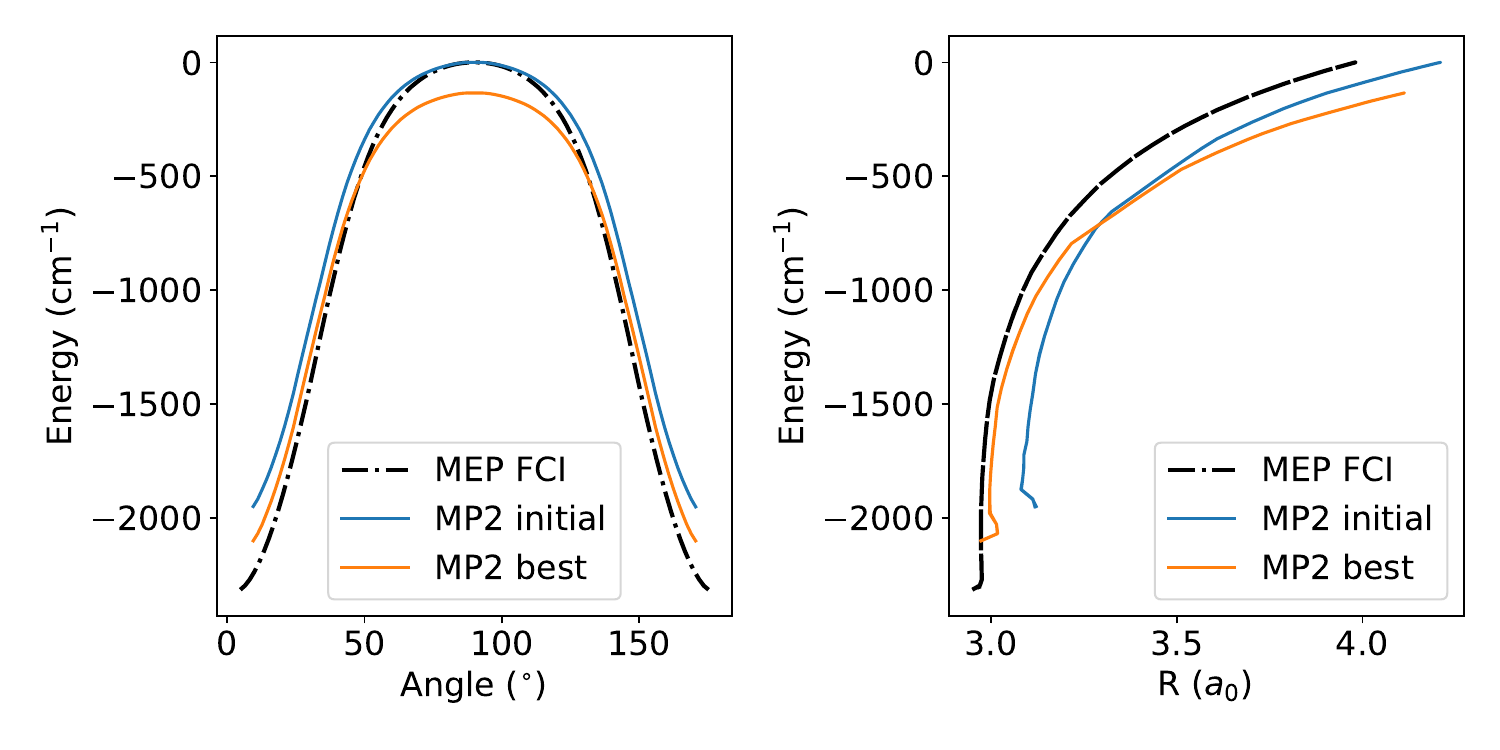}
    \caption{Minimum energy paths (MEPs) for the morph between the FCI
      and MP2 surface (PES to PES morphing) with respect to the
      variables $\theta$ and $R$. In black, the MEP for the FCI
      surface used as reference. The zero of energy of the path was
      chosen as the energy of the transition state for the unmorphed
      potential.}
    \label{sifig:meps_pes2pes}
\end{figure}

\begin{figure}
    \centering
    \includegraphics[width=\textwidth]{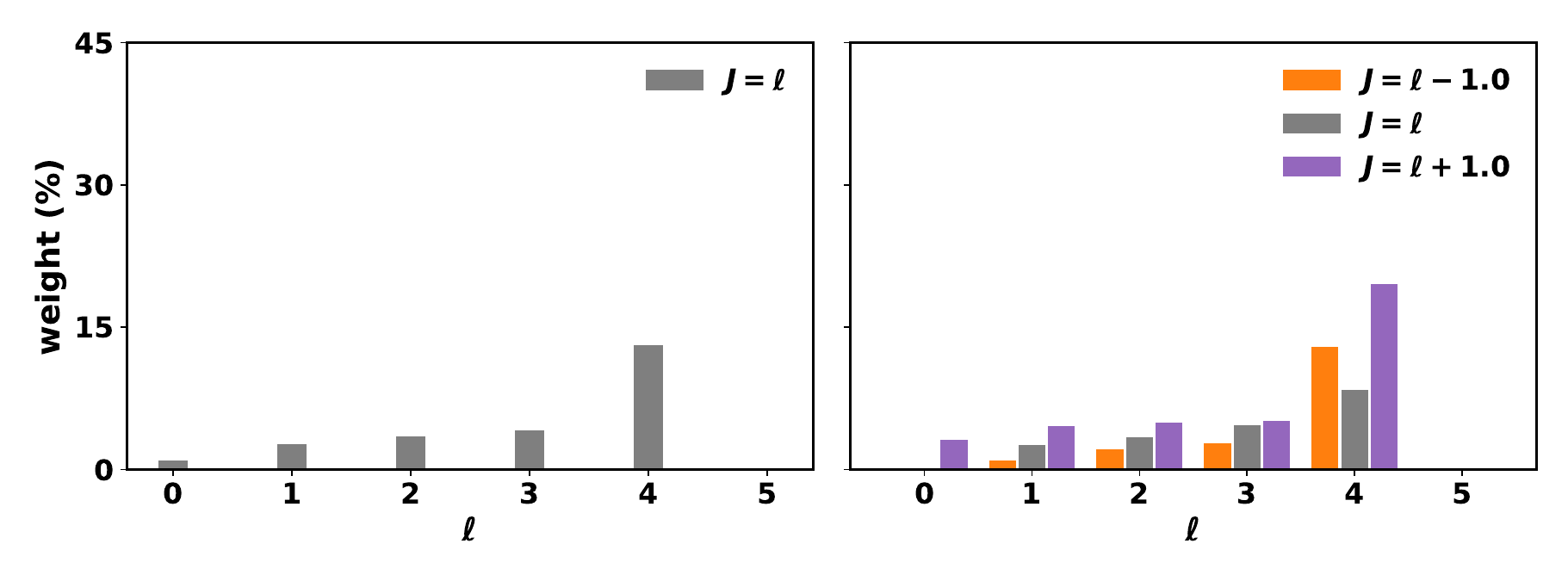}
      \caption{Relative contribution of each angular momentum state
        for collision of He($^3$P) with para- and ortho-H$_2$ (left
        and right panels, respectively) for the experimental collision
        energy $E_{\mathrm{col}}/\mathrm{k_B} \approx 2.5 \mathrm{K}$
        and spread. Note that contributions for $\ell > 5$ are not
        shown since they are negligibly small.}
      \label{sifig:weights}
\end{figure}

\begin{figure}
    \centering
    \includegraphics[width=\textwidth]{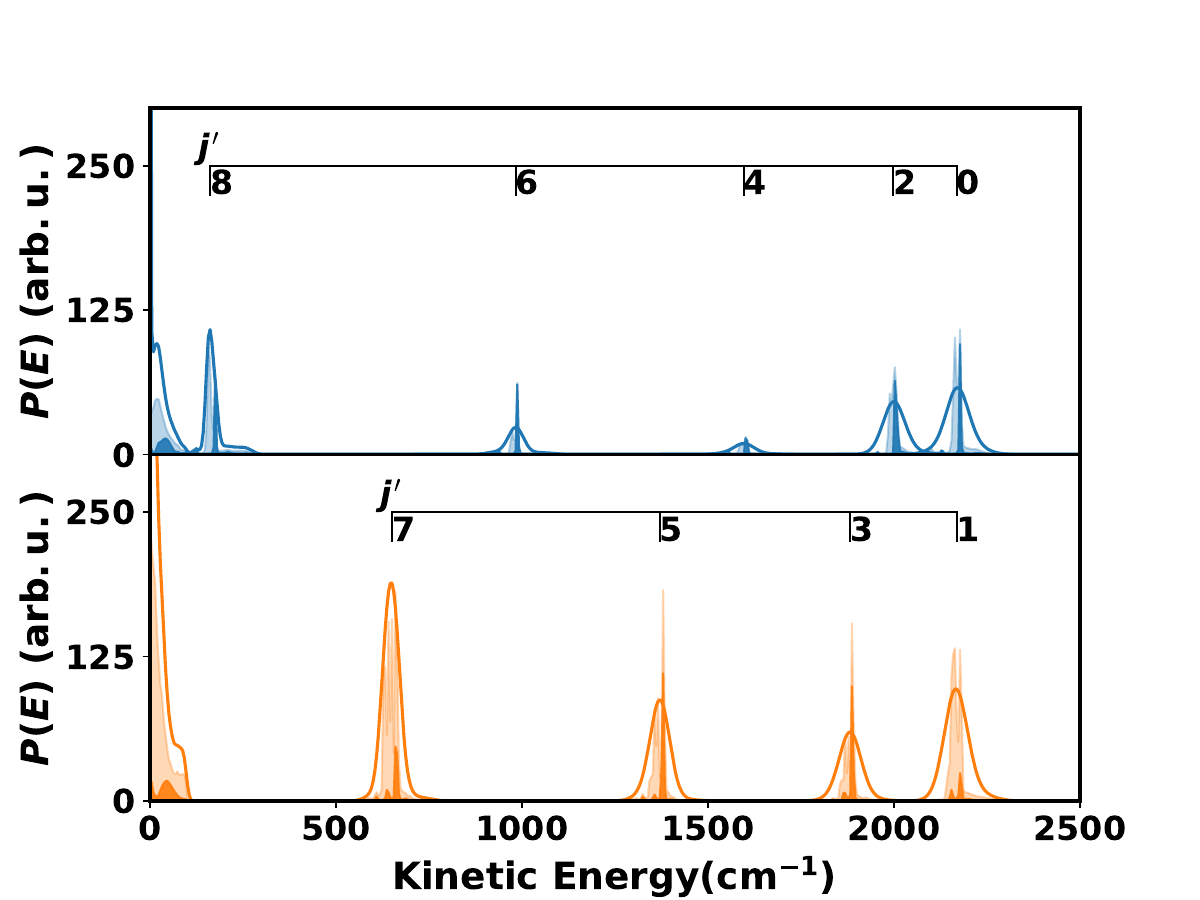} \caption{Theoretical
      kinetic energy histograms for para (above) and ortho (below)
      H$_2^+$ in their convoluted (line) and unconvoluted (shaded
      area) forms. The unconvoluted cross-sections have been scaled by
      a factor of $0.5$ relative to the convoluted ones. The darkly
      shaded areas correspond to the dominant contributions to the
      cross-section ($\ell=4, J=4$ and $\ell=4, J=5$ for para and
      ortho respectively), whilst the lightly shaded areas include all
      initial $\ell$ and $J$ contributions. The shown cross-sections
      were obtained using the unmorphed FCI potential.}
    \label{sifig:vwp1theo}
\end{figure}

\end{document}